\documentclass[aps,prb,reprint,superscriptaddress]{revtex4-2}
\usepackage{amssymb}
\usepackage{amsfonts}
\usepackage{amsmath}
\usepackage{bm}
\usepackage{braket}
\usepackage{xcolor}
\usepackage{graphicx}

\begin{document}

\title{Single-site dissipation stabilizes a superconducting nonequilibrium
steady state in a strongly correlated system}
\author{X. Z. Zhang}
\email{zhangxz@tjnu.edu.cn}
\affiliation{College of Physics and Materials Science, Tianjin Normal University, Tianjin
300387, China}

\begin{abstract}
Can superconducting order be engineered as a robust attractor of open-system
dynamics in strongly correlated systems? We demonstrate this possibility by
proposing a minimal dissipation-engineering protocol for the particle-hole
symmetric Hubbard model. By applying a rotated quantum jump operator,
specifically a locally transformed $\eta$-pair lowering operator, on a
single lattice site only, we show that the Lindblad evolution autonomously
pumps the system from the vacuum into a nonequilibrium steady state (NESS)
exhibiting macroscopic $\eta$-pair off-diagonal long-range order (ODLRO).
Crucially, this local-to-global synchronization stands in stark contrast to
schemes reliant on spatially extensive reservoirs: here, a single local
dissipative seed suffices to establish long-range coherence across the
entire interacting lattice system. We elucidate the underlying mechanism via
three core features: local dark-state selection, the controlled elimination
of off-manifold excursions induced by hopping, and a Liouvillian
invariant-subspace structure that yields an attractive fixed point with a
finite dissipative gap. Furthermore, we systematically classify the
stability of this NESS with respect to static disorder, and identify a wide
regime in which the superconducting attractor remains robust against
Hamiltonian perturbations that preserve the effective subspace structure. We
also pinpoint specific perturbations that directly cause dephasing of the $%
\eta$-pseudospin coherence and suppress ODLRO. These findings open up a
disorder-tolerant pathway for stabilizing superconducting order as a
non-thermal attractor through minimal local quantum-jump control.
\end{abstract}

\maketitle

\section{Introduction}

\label{sec:intro}

The paradigm of quantum state engineering has shifted fundamentally in
recent years: rather than treating dissipation solely as a mechanism of
decoherence, it is increasingly viewed as a resource for autonomous state
preparation and stabilization\cite%
{Lindblad1976,Gorini1976,Breuer2002,Dalibard1992,Poyatos1996,Kraus2008,Verstraete2009,Muller2012,Daley2014}%
. By engineering the coupling between a system and its environment, one can
design Liouvillian dynamics whose attractor (dark state) is a desired
quantum many-body state\cite%
{Sieberer2016,Sieberer2025,Minganti2018,Zhang2024}. This approach has
successfully demonstrated the stabilization of entangled states\cite%
{Kastoryano2011,Barreiro2011,Shankar2013}, topological phases\cite%
{Diehl2011NatPhys,Bardyn2012,Iemini2016,Goldstein2019}, and superfluid
correlations in various atomic, molecular, and optical (AMO) and solid-state
platforms\cite{Diehl2008NatPhys,Diehl2010PRL,CarusottoCiuti2013}.

However, a major open challenge lies in extending these protocols to
strongly correlated systems where competing orders and constraints make the
phase diagram fragile. Most existing schemes for stabilizing long-range
order rely on spatially extensive dissipation---engineering reservoirs that
couple to every site or bond of the lattice\cite%
{Diehl2008NatPhys,Verstraete2009}. While theoretically powerful, such bulk
engineering is experimentally demanding and often incompatible with the
natural local addressing capabilities of quantum devices. This raises a
fundamental question of control and scalability: Is it possible to stabilize
a macroscopic, strongly correlated quantum phase using only a minimal,
strictly local dissipative seed?

In this work, we answer this question in the affirmative by demonstrating
the dissipative stabilization of off-diagonal long-range order (ODLRO) in
the Hubbard model using a single local dissipator. We focus on the
paradigmatic setting of $\eta$-pairing superconductivity\cite{Yang1989Eta},
an exact eigenstate phenomenon in the Hubbard model that implies
distance-independent pairing correlations protected by an SU(2) pseudospin
symmetry. While $\eta$-pairing states are typically high-energy excited
states in equilibrium and are thus thermodynamically unstable\cite%
{Tindall2019,Mark2020},they represent ideal candidates for non-equilibrium
steady states (NESS) in the presence of appropriate dynamical symmetries 
\cite{Bernier2018,Buca2019}.

We introduce a protocol based on a rotated quantum jump operator acting on
as few as a single lattice site. Our results reveal a striking
local-to-global mechanism: this local dissipative seed effectively pumps the
entire interacting many-body system into a specific holon--doublon manifold,
triggering a self-organization process that establishes global
superconducting coherence. Crucially, this mechanism bypasses the need for
bulk reservoir engineering. Instead, the jump operator's dark-state
structure combines with coherent Hamiltonian dynamics to propagate the
locally injected pseudospin coherence across the lattice.

Our analysis bridges the gap between the formal study of open-system
symmetries and practical state preparation. Recent theoretical advances have
clarified the role of strong and weak symmetries in determining the
uniqueness and relaxation timescales of the NESS\cite%
{AlbertJiang2014,Buca2012}. Building on this, we show that our protocol
leverages a dynamical selection rule that decouples the target
superconducting manifold from competing dissipative decay channels.
Furthermore, we address the critical issue of experimental realism by
systematically investigating disorder. We identify a robust operational
regime where the superconducting NESS and its ODLRO survive against a broad
class of Hamiltonian perturbations, including static potential and
interaction disorder, which are ubiquitous in solid-state and cold-atom
experiments\cite{Znidaric2013,Beaulieu2025}.

The rest of the paper is organized as follows. In Sec.~\ref%
{sec:single_spin_heuristic}, we provide a heuristic derivation that distills
the main idea of the rotated-jump scheme and its local dark-state structure.
In Sec.~\ref{sec:eta_ketbra_section}, we present the full protocol for the
Hubbard chain and analyze the underlying mechanism from the viewpoint of
symmetry, with emphasis on the Liouvillian structure and invariant
subspaces. Section~\ref{sec:odlro_diagnostics} is devoted to the dynamical
diagnostics of superconducting behavior: we numerically track the time
evolution of two pairing correlators and demonstrate the emergence of $\eta$%
-pair ODLRO in the NESS. In Sec.~\ref{sec:disorder}, we investigate disorder
effects in a systematic manner and delineate robust versus destructive
perturbations. Finally, Sec.~\ref{sec: conclusion} summarizes our findings
and outlines future directions. Technical derivations and additional details
are provided in the Appendices.

\begin{figure}[t]
\centering
\includegraphics[width=\linewidth]{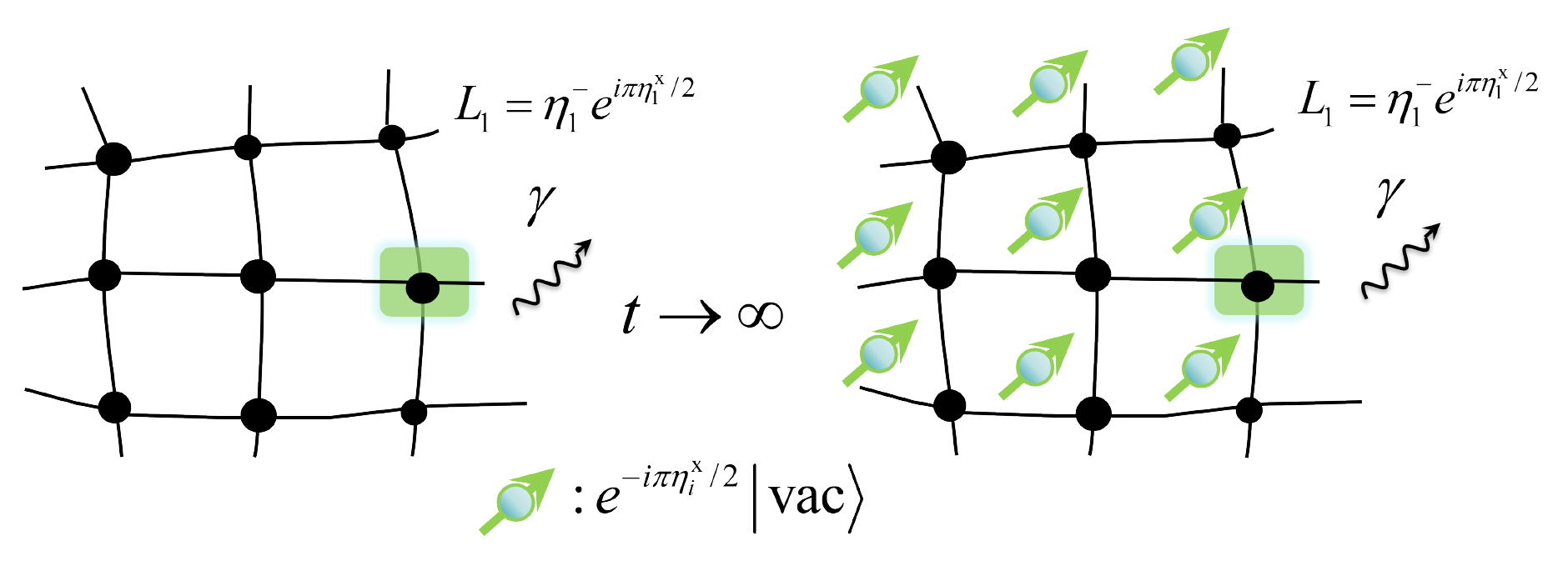}
\caption{ Schematic of the non-equilibrium phase-locking protocol that
stabilizes an $\protect\eta$-paired superconducting NESS. Starting from a
strongly correlated Hubbard system (with optional disorder in the
interaction $U_i$, onsite potential $\protect\mu_i$, and/or bond hopping $%
t_{ij}$), we apply local rotated dissipation on selected sites with jump
operators $L_j=\protect\eta_j^- e^{i\frac{\protect\pi}{2}\protect\eta_j^x}$.
Unlike the unrotated channel $\tilde L_j=\protect\eta_j^-$, which simply
annihilates pairs and leaves the vacuum as the local dark state, the
rotation tilts the dissipative measurement axis so that the locally
protected state is the $+y$ $\protect\eta$-pseudospin eigenstate, $|\protect%
\eta_j^{y}=+1/2\rangle \propto e^{-i\frac{\protect\pi}{2}\protect\eta%
_j^x}|0\rangle_j$, i.e., a coherent holon--doublon superposition. Each
quantum jump then acts like a small compass correction: it damps misaligned
components while repumping transverse $\protect\eta$ coherence. In the
presence of strong correlations, this local axis selection propagates
through the many-body dynamics, effectively phase-locking the $\protect\eta$
pseudospins across the lattice. At long times, the system relaxes to a
unique NESS with $\protect\eta$ pseudospins aligned along the $+y$
direction, yielding a finite $\protect\eta$-pair amplitude and ODLRO.}
\label{fig:illu}
\end{figure}

\section{Single-qubit heuristic: rotated dissipation pumps coherence from
the vacuum}

\label{sec:single_spin_heuristic}

We begin with the minimal open-system setting: a single spin-$1/2$ evolving
under a Lindblad master equation 
\begin{equation}
\dot{\rho}=\mathcal{L}\rho =-i[H,\rho ]+L\rho L^{\dagger }-\frac{1}{2}%
\{L^{\dagger }L,\rho \},  \label{eq:lindblad}
\end{equation}
and set $H=0$ throughout this section. The goal of this toy example is to
make one mechanism completely transparent, in a form that will reappear in
the Hubbard chain: a jump operator does not merely relax the system, it
selects an attractive fixed point (a dark state or dark manifold).
Consequently, rotating the jump operator rotates the dissipatively selected
axis and thus the steady state itself. In this way, dissipation acts not as
a destroyer of phase information, but rather as a simple coherence pump.

We first choose the standard (unrotated) lowering jump 
\begin{equation}
L_{0}=\sqrt{\gamma }\,s^{-},\qquad s^{-}=s^{x}-is^{y}.
\label{eq:jump_unrotated}
\end{equation}
Physically, this is the textbook amplitude-damping channel: population is
irreversibly transferred from $|\uparrow\rangle$ to $|\downarrow\rangle$.
Accordingly, $|\downarrow \rangle $ is a dark state since $L_{0}|\downarrow
\rangle =0$, and thus $\mathcal{L}(|\downarrow \rangle \langle \downarrow
|)=0$. Any initial state is driven toward the south pole of the Bloch
sphere; in particular, the unique steady state has no transverse coherence.

To connect with the many-body construction below, it is useful to regard
this spin-$1/2$ as a local pseudospin (the $\eta $-spin) on a Hubbard site,
with the identification 
\begin{equation}
|\downarrow \rangle \ \leftrightarrow \ |0\rangle ,\qquad |\uparrow \rangle
\ \leftrightarrow \ |\uparrow \downarrow \rangle .  \label{eq:eta_spin_map}
\end{equation}%
In this language, a transverse pseudospin polarization corresponds to a
coherent superposition of the empty site $|0\rangle $ and the doublon $%
|\uparrow \downarrow \rangle $, and therefore to a nonzero on-site pairing
amplitude $\langle \eta _{i}^{+}\rangle \neq 0$. The key question is then:
can a purely dissipative operation select a steady state with such
transverse coherence, starting from the local vacuum $|0\rangle $?

A minimal way is to rotate the jump. We consider 
\begin{equation}
L(\theta )=\sqrt{\gamma }\,s^{-}e^{i\theta s^{x}},  \label{eq:jump_rotated}
\end{equation}%
which can be viewed as applying the usual damping $s^{-}$ in a locally
rotated basis. Equivalently, the dissipator is engineered so that the dark
state is no longer the south pole, but a state at a tunable latitude on the
Bloch sphere. The special choice $\theta =\pi /2$ rotates the attractor onto
the equator and thereby maximizes the transverse coherence. For a single
jump channel, a pure steady state is most transparently obtained from the
dark-state condition $L(\theta )|\psi _{\mathrm{d}}\rangle =0$. For $\theta
=\pi /2$, $L_{\pi /2}|\psi _{\mathrm{d}}\rangle =0\Longleftrightarrow
s^{-}\!\left( e^{i\frac{\pi }{2}s^{x}}|\psi _{\mathrm{d}}\rangle \right)
=0\Longleftrightarrow e^{i\frac{\pi }{2}s^{x}}|\psi _{\mathrm{d}}\rangle
\propto |\downarrow \rangle $, and hence (see Appendix~\ref%
{app:single_qubit_liouvillian} for more details) 
\begin{equation}
|\psi _{\mathrm{d}}\rangle \propto e^{-i\frac{\pi }{2}s^{x}}|\downarrow
\rangle =\frac{1}{\sqrt{2}}\left( |\downarrow \rangle -i|\uparrow \rangle
\right) .  \label{eq:dark_state_rotated}
\end{equation}%
This is the central point: the steady state is still a dark state of the
jump, but it now lies on the equator of the Bloch sphere and therefore
carries a finite transverse coherence, 
\begin{equation}
\langle \psi _{\mathrm{d}}|\,s^{+}\,|\psi _{\mathrm{d}}\rangle =\frac{i}{2}%
\neq 0.  \label{eq:coherence_single_spin}
\end{equation}%
Equivalently, the steady-state Bloch vector points along $+\hat{y}$, with $%
\langle s^{x}\rangle =0$ and $\langle s^{y}\rangle =1/2$. Under the mapping %
\eqref{eq:eta_spin_map}, Eq.~\eqref{eq:coherence_single_spin} is precisely
the single-site analogue of $\langle \eta _{i}^{+}\rangle \neq 0$: the
rotated dissipation does not create coherence out of nothing; rather, it
reshapes the dark-state structure so that the unique attractor contains the
desired coherence between $|0\rangle $ and $|\uparrow \downarrow \rangle $.

This local picture foreshadows the many-body mechanism exploited below. In
the Hubbard chain, implementing the analogous rotated jump in the $\eta $%
-spin sector selects locally coherent holon--doublon superpositions as
building blocks. The remaining task is to understand how interactions and
hopping propagate and phase-lock this locally seeded $\eta $-pseudospin
orientation across the lattice, resulting in an $\eta $-paired NESS with
off-diagonal long-range order.

\section{Symmetry structure, holon--doublon protection, and the Liouvillian
spectrum}

\label{sec:eta_ketbra_section}

The single-qubit example shows that a rotated jump operator can reshape the
local dark state and pump coherence along a chosen pseudospin direction. In
the Hubbard setting, the same idea becomes nontrivial for two reasons: (i)
the jump operators are local and therefore need not preserve global SU(2)
quantum numbers (in particular the global operator $\eta ^{2}$), and (ii)
the hopping term couples the holon--doublon sector to intermediate
singly-occupied configurations. In this section we identify the dynamically
relevant manifold, clarify which symmetries remain operative in the open
system, and state the effective projected Liouvillian governing the
long-time relaxation toward an $\eta $-paired NESS. A detailed derivation of
the effective generator is provided in Appendix~\ref{app:eff_L_derivation}.

\subsection{Particle--hole symmetric Hubbard model and $\protect\eta$-SU(2)}

We consider the particle--hole symmetric Hubbard model on a bipartite
lattice, 
\begin{eqnarray}
H_{\mathrm{Hub}} &=&-t\sum_{\langle ij\rangle ,\sigma }\!\left( c_{i\sigma
}^{\dagger }c_{j\sigma }+\mathrm{H.c.}\right)  \notag \\
&&+U\sum_{i}\left( n_{i\uparrow }-\frac{1}{2}\right) \left( n_{i\downarrow }-%
\frac{1}{2}\right) ,  \label{eq:Hubbard_phs_sec}
\end{eqnarray}%
where half filling is enforced by the particle--hole symmetric interaction
term. The model has the global SO(4) symmetry $\mathrm{SO}(4)\simeq \mathrm{%
SU}(2)_{S}\times \mathrm{SU}(2)_{\eta }/\mathbb{Z}_{2}$. Since our protocol
targets pairing coherence, we focus on the $\eta $-pseudospin generators 
\begin{eqnarray}
\eta ^{\pm } &=&\sum_{i}\eta _{i}^{\pm },\qquad \\
\eta _{i}^{+} &=&(-1)^{i}c_{i\uparrow }^{\dagger }c_{i\downarrow }^{\dagger
},\qquad \\
\eta _{i}^{z} &=&\frac{1}{2}\left( n_{i\uparrow }+n_{i\downarrow }-1\right) ,
\label{eq:eta_generators_sec}
\end{eqnarray}%
with $\eta _{i}^{-}=(\eta _{i}^{+})^{\dagger }$ and $\eta _{i}^{x,y}$
defined in the standard way. They satisfy $[\eta _{i}^{\alpha },\eta
_{j}^{\beta }]=i\delta _{ij}\epsilon _{\alpha \beta \gamma }\eta
_{i}^{\gamma }$. In the particle--hole symmetric form %
\eqref{eq:Hubbard_phs_sec}, one has the strong commutation relations 
\begin{equation}
\lbrack H_{\mathrm{Hub}},\eta ^{\pm }]=0,\qquad \lbrack H_{\mathrm{Hub}%
},\eta ^{z}]=0,  \label{eq:H_comm_eta_pm}
\end{equation}%
and hence also $[H_{\mathrm{Hub}},\eta ^{2}]=0$.

\subsection{Local rotated dissipation and the relevant invariant structure}

We introduce rotated local jump operators acting on an arbitrary subset of
sites $\mathcal{J}\subseteq \{1,\dots ,N\}$, 
\begin{eqnarray}
L_{j} &=&\sqrt{\gamma }\,\eta _{j}^{-}e^{i\frac{\pi }{2}\eta _{j}^{x}}=\sqrt{%
\gamma }\,U_{j}^{\dagger }\eta _{j}^{-}U_{j},\qquad
\label{eq:eta_jump_local_sec} \\
U_{j} &=&e^{-i\frac{\pi }{2}\eta _{j}^{x}},\qquad j\in \mathcal{J}.
\end{eqnarray}%
The Lindblad evolution is 
\begin{equation}
\partial _{t}\rho =\mathcal{L}\rho =-i[H_{\mathrm{Hub}},\rho ]+\sum_{j\in 
\mathcal{J}}\left( L_{j}\rho L_{j}^{\dagger }-\frac{1}{2}\{L_{j}^{\dagger
}L_{j},\rho \}\right) .  \label{eq:lindblad_master}
\end{equation}

A key distinction from collective dissipation is that local $L_{j}$
generally do not commute with the global operator $\eta ^{2}=(\sum_{i}\vec{%
\eta}_{i})^{2}$ because $\eta ^{2}$ contains cross-terms $\sum_{k\neq \ell }%
\vec{\eta}_{k}\!\cdot \!\vec{\eta}_{\ell }$. Thus, in general, 
\begin{equation}
\lbrack L_{j},\eta ^{2}]\neq 0,  \label{eq:Lj_not_comm_eta2}
\end{equation}%
and the open dynamics is not block-diagonal in global $\eta $ sectors. This
point is crucial for internal consistency: the $(N+1)$-dimensional $\eta $%
-multiplet generated from the vacuum, 
\begin{equation}
|\Psi _{M}\rangle \propto (\eta ^{+})^{M}|\mathrm{vac}\rangle ,\qquad
M=0,1,\dots ,N,  \label{eq:PsiM_def}
\end{equation}%
is therefore not an exact invariant subspace of the full local-dissipative
dynamics.

What is protected microscopically is the local representation structure of $%
\vec{\eta}_{i}$. The on-site Hilbert space decomposes into a singlet
(singly-occupied) sector $\{|\uparrow \rangle _{i},|\downarrow \rangle
_{i}\} $ with $\eta _{i}=0$ and a doublet (holon--doublon) sector $%
\{|0\rangle _{i},|\uparrow \downarrow \rangle _{i}\}$ with $\eta _{i}=\frac{1%
}{2}$. The local $\eta _{i}^{\alpha }$ act as spin-$1/2$ generators only
within the doublet and annihilate the singlet. In particular, 
\begin{equation}
\eta _{i}^{\alpha }|\uparrow \rangle _{i}=\eta _{i}^{\alpha }|\downarrow
\rangle _{i}=0,\qquad L_{i}|\uparrow \rangle _{i}=L_{i}|\downarrow \rangle
_{i}=0.  \label{eq:eta_annih_singlons}
\end{equation}%
Consequently, the dissipators themselves never create singlons. Starting
from the fermionic vacuum $|\mathrm{vac}\rangle =\bigotimes_{i}|0\rangle
_{i} $ (all sites in the $\eta _{i}=\frac{1}{2}$ doublet), the dissipative
channel acts entirely within the tensor-product holon--doublon space 
\begin{equation}
\mathcal{H}_{\mathrm{HD}}=\bigotimes_{i=1}^{N}\mathrm{span}\{|0\rangle
_{i},|\uparrow \downarrow \rangle _{i}\},\qquad \dim \mathcal{H}_{\mathrm{HD}%
}=2^{N}.  \label{eq:HD_space}
\end{equation}%
However, the hopping term $H_{t}$ does connect holon/doublon configurations
to intermediate singly-occupied states and hence can generate virtual (and,
away from strict limits, small but finite) excursions out of $\mathcal{H}_{%
\mathrm{HD}}$.

\subsection{Why the coherent generator drops out in the $\protect\eta$%
-multiplet}

Although $\mathcal{H}_{\Psi }=\mathrm{span}\{|\Psi _{M}\rangle \}$ is not
strictly invariant under local dissipation, it remains the natural slow
sector for constructing an effective description because it is an exactly
degenerate manifold of the particle--hole symmetric Hubbard Hamiltonian.
Using \eqref{eq:H_comm_eta_pm} and the fact that the vacuum has zero energy
in the particle--hole symmetric convention, 
\begin{equation}
H_{\mathrm{Hub}}|\mathrm{vac}\rangle =0,  \label{eq:vac_energy_zero}
\end{equation}%
one obtains for all $M$, 
\begin{equation}
H_{\mathrm{Hub}}|\Psi _{M}\rangle =H_{\mathrm{Hub}}(\eta ^{+})^{M}|\mathrm{%
vac}\rangle =(\eta ^{+})^{M}H_{\mathrm{Hub}}|\mathrm{vac}\rangle =0.
\label{eq:H_on_PsiM_zero}
\end{equation}%
Therefore, for any operator $\rho $ supported on $\mathcal{H}_{\Psi }$ one
has 
\begin{equation}
-i[H_{\mathrm{Hub}},\rho ]=0.  \label{eq:commutator_vanish_on_HPsi}
\end{equation}%
Equivalently, in the doubled (ket$\otimes $bra) representation, the coherent
contribution $-i(H_{\mathrm{Hub}}\otimes \mathbb{I}-\mathbb{I}\otimes H_{%
\mathrm{Hub}}^{T})$ vanishes identically on $\mathcal{H}_{\Psi }\otimes 
\mathcal{H}_{\Psi }^{\ast }$. This observation motivates a degenerate
projected effective theory on the doubled $\eta $-multiplet: the leading
generator is obtained by projecting the dissipator onto $\mathcal{H}_{\Psi
}\otimes \mathcal{H}_{\Psi }^{\ast }$. Because the local jumps do not
conserve $\eta ^{2}$, this projection is not exact; rather, it is the first
term in a controlled elimination of off-manifold components, in which both
(i) intra-holon--doublon mixing among different global-$\eta $ sectors
induced by $L_{j}$ and (ii) virtual leakage to singlons induced by $H_{t}$
generate higher-order corrections. The resulting picture is fully consistent
with the numerics: the attractive NESS is fixed already at leading order,
while higher orders mainly renormalize transient relaxation pathways, which
will show in the next section.

\subsection{Leading projected Liouvillian, NESS, and the dissipative gap}

Projecting the dissipator onto the doubled $\eta $-multiplet and using the $%
\eta $-SU(2) algebra yields the following leading effective generator
(derived in Appendix~\ref{app:eff_L_derivation}): 
\begin{equation}
\hat{\mathcal{L}}_{\mathrm{eff}}^{(0)}=-\frac{\gamma |\mathcal{J}|}{2}%
+\gamma \sum_{j\in \mathcal{J}}\left[ \frac{1}{2}\left( \eta _{j}^{y}-\tilde{%
\eta}_{j}^{y}\right) +\eta _{j}^{-}e^{i\frac{\pi }{2}\eta _{j}^{x}}\,\tilde{%
\eta}_{j}^{-}e^{-i\frac{\pi }{2}\tilde{\eta}_{j}^{x}}\right] ,
\label{eq:exact_liouvillian_form}
\end{equation}%
where $\eta _{j}$ ($\tilde{\eta}_{j}$) acts on the ket (bra) space.
Importantly, Eq.~\eqref{eq:exact_liouvillian_form} should be understood as a
degenerate first-order projected Liouvillian on $\mathcal{H}_{\Psi }\otimes 
\mathcal{H}_{\Psi }^{\ast }$, not as the full microscopic generator for
arbitrary parameters.

The steady state follows immediately. The rotated jump annihilates the local 
$+y$ eigenstate, 
\begin{equation}
\eta _{j}^{-}e^{i\frac{\pi }{2}\eta _{j}^{x}}\,|\eta _{j}^{y}=+1/2\rangle =0,
\label{eq:local_dark_y}
\end{equation}%
and therefore the product density matrix 
\begin{equation}
\rho _{\mathrm{ss}}=\bigotimes_{j=1}^{N}|\eta _{j}^{y}=+1/2\rangle \langle
\eta _{j}^{y}=+1/2|  \label{eq:steady_state_product}
\end{equation}%
is a simultaneous dark state for any nonempty driven set $\mathcal{J}$. In
particular, 
\begin{equation}
|\langle \eta _{i}^{+}\rangle _{\mathrm{ss}}|=\frac{1}{2},\qquad |\langle
\eta _{i}^{+}\eta _{j}^{-}\rangle _{\mathrm{ss}}|=\frac{1}{4}\quad (i\neq j),
\label{eq:ss_values}
\end{equation}%
consistent with the ODLRO of an $\eta $-paired condensate.

We finally address the relaxation timescale and robustness. The operator in %
\eqref{eq:exact_liouvillian_form} has a triangular structure in the $\eta
^{y}$ basis (Appendix~\ref{app:triangular_proof}), implying that the
eigenvalues are given by the diagonal entries and that the slowest nonzero
decay rate is set by a single local $y$-pseudospin flip. This yields a
leading dissipative gap 
\begin{equation}
\Delta =\frac{\gamma }{2}.  \label{eq:gap_main}
\end{equation}%
The key physical point is that this gap is already present at the level of
the projected generator, making $\rho _{\mathrm{ss}}$ a stable attractive
fixed point. Deviations from the projected description arise from two
distinct mechanisms: (i) intra-holon--doublon excursions in which local $%
L_{j}$ mixes different global-$\eta $ sectors within $\mathcal{H}_{\mathrm{HD%
}}$, and (ii) leakage into the singlon sector mediated primarily by the
hopping $H_{t}$. Both effects enter perturbatively through higher-order
elimination of the off-manifold components and, in the parameter regime
explored numerically, renormalize transient relaxation pathways without
altering the observed NESS \eqref{eq:steady_state_product} or the dominant
decay scale set by $\gamma /2$.


To address potential objections, we offer three key clarifications. (i)\ 
\emph{Effective gap and timescales:} Eq.~\eqref{eq:gap_main} describes a
property of the projected effective Liouvillian--specifically, that the
leading-order reduced generator exhibits a finite spectral separation scale
set by the local pumping strength. In the full microscopic model, additional
processes (e.g., hopping-assisted excursions and local sector mixing)
renormalize relaxation pathways and hinder the clean extraction of a
single-exponential rate. Rather than overinterpreting fitted microscopic gap
values, we adopt $\Delta \sim \gamma /2$ as an organizing scale that
accounts for the rapid convergence and suppressed deviations within the
strong-coupling/controlled regime where effective elimination holds. (ii)%
\emph{Finite-size scaling:} To distinguish genuine long-range order from
finite-size saturation, we track the $\eta $-pair structure factor $S_{\eta
}(\tau )$ as well as the global amplitude $|\Phi (\tau )|$ for progressively
larger system sizes, with results presented in the next subsection. The weak
size dependence exhibited by these steady-state plateaus confirms an $%
\mathcal{O}(1)$ order parameter in the thermodynamic limit. (iii)\ \emph{%
Uniqueness within the studied sector:} Starting from the vacuum
(holon--doublon) sector, the rotated jumps yield a unique attractive fixed
point in our simulations. In practice, the key question is not whether the projected manifold is exactly microscopically invariant, but whether off-manifold processes induce measurable steady-state deviations in accessible system sizes and times, and in the regime we consider, such deviations are absent.
\section{Dynamics: buildup of ODLRO and convergence to an $\protect\eta$%
-paired NESS}

\label{sec:odlro_diagnostics}

The key claim of this work is that rotated local dissipation stabilizes a
non-equilibrium superconducting steady state emerging from the fermionic
vacuum, 
\begin{equation}
|\mathrm{vac}\rangle \equiv \bigotimes_{i=1}^{N}|0\rangle _{i},
\end{equation}%
specifically as defined by ODLRO in the $\eta $-pair channel. This choice of
initial state is essential: $|\mathrm{vac}\rangle $ lies entirely in the
holon--doublon doublet on every site and therefore provides a clean entry
point for the dissipative pumping mechanism analyzed in Sec.~\ref%
{sec:single_spin_heuristic} and the symmetry-based reduction of Sec.~\ref%
{sec:eta_ketbra_section}. In particular, the protocol does not rely on
fine-tuning a preformed paired state; rather, coherence is built up
dynamically by the local jumps, which select a transverse $\eta $-pseudospin
axis and actively phase-lock the system into an $\eta $-paired condensate.

To make this statement quantitative, we track two standard diagnostics of
pairing coherence: (i) the one-point $\eta$-pair amplitude, which measures
on-site holon--doublon coherence, and (ii) the real-space $\eta$-pair
correlator as a function of separation, whose long-distance plateau
constitutes the defining signature of ODLRO. Together, these observables
provide a direct bridge between the single-spin pumping picture of Sec.~\ref%
{sec:single_spin_heuristic} and the projected-Liouvillian analysis of Sec.~%
\ref{sec:eta_ketbra_section}, and they allow us to verify explicitly that
the long-time NESS exhibits superconducting order in the $\eta$-pair channel.

We examine pairing coherence using the local $\eta $-pair operator $\eta
_{i}^{+}$ (and its Hermitian conjugate $\eta _{i}^{-}$). The local pairing
amplitude, 
\begin{equation}
\Phi _{i}(\tau )\equiv \langle \eta _{i}^{+}(\tau )\rangle ,\qquad \Phi
(t)\equiv \frac{1}{N}\sum_{i=1}^{N}\langle \eta _{i}^{+}(\tau )\rangle ,
\label{eq:Phi_def_odlro}
\end{equation}%
directly measures the on-site coherence between $|0\rangle _{i}$ and $%
|\uparrow \downarrow \rangle _{i}$ (cf.~the pseudospin mapping in Eq.~%
\eqref{eq:eta_spin_map}). To establish ODLRO we consider the two-point
correlator 
\begin{equation}
C_{ij}(\tau )\equiv \langle \eta _{i}^{+}(\tau )\eta _{j}^{-}(\tau )\rangle ,
\label{eq:Cij_def_odlro}
\end{equation}%
and, for periodic systems, its translationally averaged version 
\begin{equation}
C(r,\tau )\equiv \frac{1}{N}\sum_{i=1}^{N}\langle \eta _{i}^{+}(\tau )\eta
_{i+r}^{-}(\tau )\rangle .  \label{eq:Cr_def_odlro}
\end{equation}%
ODLRO corresponds to a nonzero long-distance limit, 
\begin{equation}
C_{\infty }(\tau )\equiv \lim_{r\rightarrow \infty }C(r,t)>0,
\label{eq:ODLRO_condition}
\end{equation}%
or, equivalently, to a finite $\eta $-pair structure factor 
\begin{equation}
S_{\eta }(\tau )\equiv \frac{2}{N\left( N-1\right) }\sum_{i<j}\langle \eta
_{i}^{+}(t)\eta _{j}^{-}(t)\rangle ,  \label{eq:Seta_def}
\end{equation}%
which remains $\mathcal{O}(1)$ in the thermodynamic limit.

\begin{figure}[t]
\centering
\includegraphics[width=%
\linewidth]{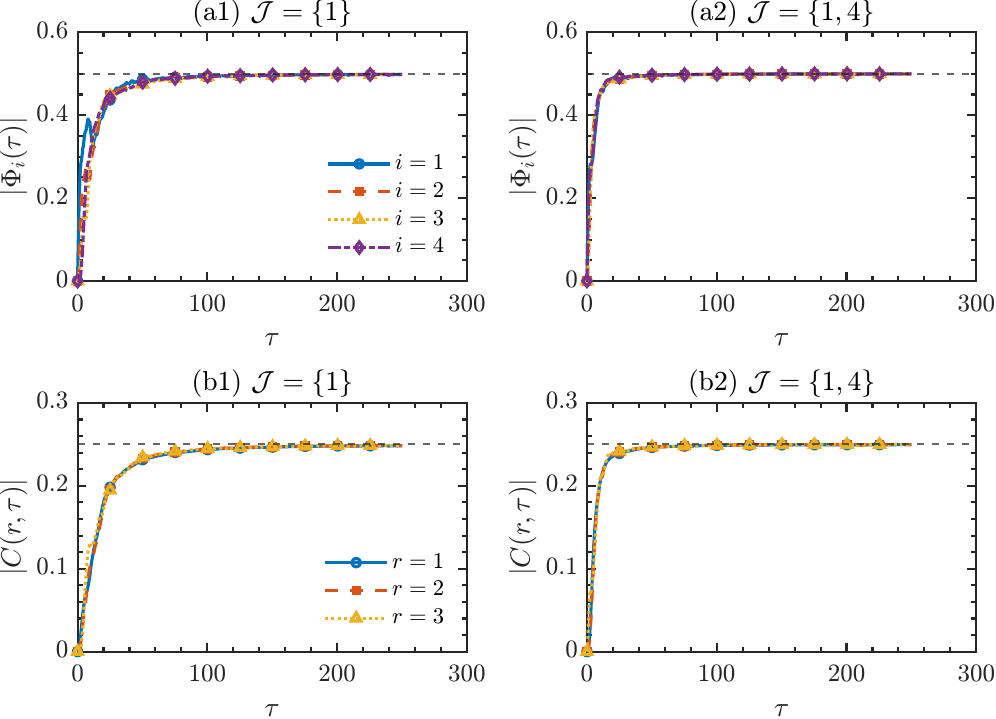}
\caption{ Real-time buildup of $\protect\eta$-pair coherence under open
boundary conditions (OBC) for $N=4$ at $t=\protect\gamma$, $U=8\gamma$, and the rotation angle $\protect\theta=\protect\pi/2$. Top panels:
site-resolved one-point amplitude $|\Phi_i(\protect\tau)|=|\langle\protect%
\eta_i^+(t)\rangle|$ for (a1) a single driven site $\mathcal{J}=\{1\}$ and
(a2) two driven sites $\mathcal{J}=\{1,4\}$; the dashed line indicates the
projected benchmark $|\Phi_i|=1/2$. Bottom panels: real-space profile of the
two-point correlator magnitude $|C(r,\protect\tau)|=|\langle \protect\eta%
_i^+(t)\protect\eta_{i+r}^-(t)\rangle|$ for multiple separations $r$ (here $%
r=1,2,3$), shown for (b1) $\mathcal{J}=\{1\}$ and (b2) $\mathcal{J}=\{1,4\}$%
; the dashed line marks the predicted ODLRO plateau $|C|=1/4$. At
intermediate times, short-range correlations develop first, while at late
times the correlator becomes nearly independent of $r$ and converges to an $r
$-independent plateau $|C(r,t\!\to\!\infty)|\simeq 1/4$ (up to finite-size
effects), providing direct evidence for ODLRO according to Eq.~
\eqref{eq:ODLRO_condition}. }
\label{fig:dynamics_OBC_N4_panels}
\end{figure}

At the level of the leading projected generator %
\eqref{eq:exact_liouvillian_form}, the NESS can be viewed equivalently as a
coherent $\eta $-spin state obtained by rotating the trivial holon vacuum ($%
M=0$) into the $+y$ direction
\begin{equation}
|\mathrm{NESS}\rangle =\bigotimes_{j=1}^{N}|\eta _{j}^{y}=+1/2\rangle
\;\propto \;\exp \!\Bigl(-i\frac{\pi }{2}\sum_{j=1}^{N}\eta _{j}^{x}\Bigr)\,|%
\mathrm{vac}\rangle ,  \label{eq:NESS_as_rotated_vacuum}
\end{equation}%
so the long-time state is not built by Hamiltonian pairing but rather
selected as a Lindbladian attractor by the rotated local jumps. This
directly gives the projected benchmark, namely an ODLRO plateau of $%
C_{\infty }=1/4$ (up to finite-size corrections). Beyond the strictly
projected picture, the full microscopic evolution admits perturbative
excursions outside the projected manifold. These arise from (i)
intra-holon--doublon mixing among different global-$\eta $ sectors induced
by local $L_{j}$, and (ii) virtual leakage to singly occupied configurations
mediated primarily by the hopping $H_{t}$. As discussed in Sec.~\ref%
{sec:eta_ketbra_section}, such effects enter as higher-order corrections in
the Schur-complement elimination of off-manifold components and
predominantly renormalize transient relaxation pathways, while the leading
attractive fixed point and the dominant decay scale $\Delta =\gamma /2$ [Eq.~%
\eqref{eq:gap_main}] remain intact in the regime explored below.

We now verify these predictions via direct Lindblad time evolution and explicit
computation of $\langle O\rangle _{\tau }=\mathrm{Tr}[\rho (\tau )O]$. Figures~%
\ref{fig:dynamics_OBC_N4_panels}(a1) and (a2) show the time evolution of both the
local $\eta $-pair amplitude $|\Phi _{i}(\tau )|\equiv |\langle \eta
_{i}^{+}\rangle |$ and a representative two-point correlator $|C(r,\tau
)| $ for multiple choices of the driven set $\mathcal{J}$. In all cases, the
dynamics exhibits a rapid approach to the projected values, with $|\Phi
_{i}(\tau \rightarrow \infty )|\rightarrow 1/2$ on a timescale governed by
the dissipative gap. More importantly, Figures~\ref%
{fig:dynamics_OBC_N4_panels}(b1) and (b2) display the full real-space
profile $|C(r,\tau )|$ for multiple separations $r$, providing a direct
visualization of the buildup of ODLRO. At intermediate times short-range
correlations develop first, whereas at late times the correlator becomes
nearly independent of $r$ and converges to an $r$-independent plateau, $%
|C(r,\tau \rightarrow \infty )|\simeq 1/4$ for all $1\leq r\leq N-1$ (up to
finite-size effects). This long-distance saturation constitutes direct
evidence for ODLRO according to Eq.~\eqref{eq:ODLRO_condition}.

To further quantify the macroscopic order, we track the global amplitude $%
|\Phi(\tau)|$ and the structure factor $|S_{\eta}(\tau)|$; see Fig.~\ref%
{fig:global_Phi_Seta_scaling}. For dissipation applied only on the first
site ($\mathcal{J}=\{1\}$), both observables rapidly approach steady
plateaus that are essentially insensitive to system size (as $N$ is
increased), indicating that the observed order is not a finite-size
artifact. In particular, the saturation of $|S_{\eta}(\tau)|$ to a finite
plateau consistent with the real-space ODLRO limit provides a complementary,
bulk-sensitive confirmation of the dissipatively stabilized $\eta$-paired
NESS.

Before ending this part, we note that the resulting superconducting order is
essentially a non-equilibrium phenomenon. The NESS does not emerge from
thermalization under $H_{\mathrm{Hub}}$: within the degenerate $\eta $%
-multiplet $\mathcal{H}_{\Psi }$, the Hamiltonian itsself does not prefer a
specific coherent orientation. Instead, the rotated local jumps act as a
dissipative axis selector and continuously phase-lock the $\eta $
pseudospins, making the ODLRO plateau an attractor of the open-system
dynamics. This picture aligns with symmetry considerations: since the local
dissipation generally breaks the global $\eta ^{2}$ [Eq.~%
\eqref{eq:Lj_not_comm_eta2}] symmetry, the steady state cannot be viewed as
an equilibrium (or generalized-Gibbs) ensemble dictated solely by the conserved
charges of $H_{\mathrm{Hub}}$. Rather, ODLRO persists because it is actively
stabilized by the driven dissipative dynamics, realizing a genuinely
non-equilibrium $\eta $-paired steady state.

\begin{figure}[t]
\centering
\includegraphics[width=\linewidth]{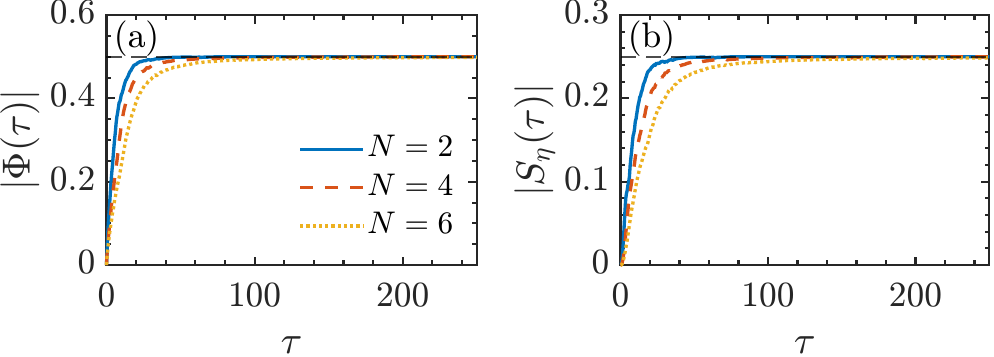}
\caption{ Finite-size scaling of global $\protect\eta$-pair order parameters
for dissipation applied only on the first site $\mathcal{J}=\{1\}$. (a)
Global $\protect\eta$-pair amplitude $|\Phi(\protect\tau)|$. (b) $\protect%
\eta$-pair structure factor $|S_{\protect\eta}(\protect\tau)|$. Curves are
shown for different system sizes (as labeled). Both observables rapidly
approach steady plateaus that are essentially insensitive to $N$, indicating
that the observed macroscopic order is not a finite-size artifact. In
particular, the saturation of $|S_{\protect\eta}(\protect\tau)|$ to a finite
value, consistent with the real-space long-distance ODLRO limit, provides a
complementary, bulk-sensitive confirmation of the dissipatively stabilized $%
\protect\eta$-paired NESS. Parameters: 1D chain, $t=\protect\gamma$, $U=4\gamma$, and $\protect\theta=\protect\pi/2$. }
\label{fig:global_Phi_Seta_scaling}
\end{figure}



\section{Disorder: Stability and Breakdown of Dissipative $\protect\eta$%
-Pair ODLRO}

\label{sec:disorder}

Having established the dynamical buildup of $\eta $-pair coherence and the
emergence of ODLRO in the clean system, we now address robustness: which
microscopic imperfections leave the dissipatively prepared $\eta $-paired
NESS essentially intact, and which disorders genuinely suppress (or destroy)
the long-distance plateau? The answer is controlled by two closely related
criteria: (i) whether the perturbation preserves the effective
holon--doublon (HD) closure $\mathcal{H}_{\mathrm{HD}}=\bigotimes_{i}\mathrm{%
span}\{|0\rangle _{i},|\uparrow \downarrow \rangle _{i}\}$ as an
(approximately) closed dynamical sector, and (ii) within that sector,
whether it competes with the dissipative pinning that selects a local
pseudospin polarization along $+y$ direction.

To quantify robustness in a way that is transparent for finite systems, we
track two steady-state diagnostics: the one-point amplitude
$|\Phi_m|_{\mathrm{ss}}$ and, more stringently, the long-distance plateau
$|C_m(r\!\to\!N/2)|_{\mathrm{ss}}$. The plateau is a direct test of ODLRO and
is typically more sensitive than local coherence to symmetry breaking and to
leakage out of the targeted manifold. In what follows, we keep the
rotated-jump channel fixed and introduce only static (quenched) perturbations
either in the Hamiltonian or in the jump parameters. The question is then
whether, after disorder averaging, the NESS remains close to the clean
$+y$-polarized fixed point for the system sizes accessible to exact
time evolution.

A useful organizing principle, consistent with Secs.~\ref{sec:eta_ketbra_section}
and \ref{sec:odlro_diagnostics}, is that the preparation protocol rests on two
distinct structures, and disorder may compromise either one:
(i) the \emph{local} holon--doublon (HD) closure
$\mathcal{H}_{\mathrm{HD}}=\bigotimes_{i}\mathrm{span}\{|0\rangle_{i},
|\uparrow\downarrow\rangle_{i}\}$, which is enforced by the local $\eta$-pseudospin
selection rules of the designed jump; and
(ii) the \emph{global} slow manifold $\mathcal{H}_{\Psi}$ (the doubled $\eta$-multiplet
supporting the triangular and gapped effective generator), within which the coherent
Hubbard contribution can be eliminated consistently to leading order.
In the clean limit these ingredients work in concert: the rotated jump pins each local
pseudospin close to the $+y$ direction, while the dissipative gap controls the late-time
relaxation into the (product-like) dark attractor.

We now add quenched disorder to the microscopic Hubbard model,
\begin{equation}
H=H_{\mathrm{Hub}}+H_{\mathrm{dis}},
\end{equation}
while keeping the same rotated local jump operators
$L_{j}=\sqrt{\gamma}\,\eta_{j}^{-}e^{i\frac{\pi}{2}\eta_{j}^{x}}$
on a chosen driven set.
Numerically, we assess robustness by averaging over $N_{\mathrm{dis}}$ disorder
realizations. For each realization, we obtain the steady state either by evolving the
Lindblad dynamics to long times or by directly solving for the zero-eigenvalue mode of
the Liouvillian, and we evaluate
\begin{align}
|\Phi_{m}|_{\mathrm{ss}} &\equiv \frac{1}{N}\,\mathbb{E}\Bigl|\sum_{i=1}^{N}
\mathrm{Tr}\!\left(\rho_{\mathrm{ss}}\,\eta_{i}^{+}\right)\Bigr|,\\
|C_{m}(r)|_{\mathrm{ss}} &\equiv \frac{1}{N}\,\mathbb{E}\Bigl|\sum_{i=1}^{N}
\mathrm{Tr}\!\left(\rho_{\mathrm{ss}}\,\eta_{i}^{+}\eta_{i+r}^{-}\right)\Bigr|,
\end{align}
where $\mathbb{E}[\cdots]$ denotes the disorder average.
In the ideal projected picture of Sec.~\ref{sec:eta_ketbra_section}, the product dark
state yields $|\langle\eta_{i}^{+}\rangle_{\mathrm{ss}}|=1/2$ and
$|\langle\eta_{i}^{+}\eta_{j}^{-}\rangle_{\mathrm{ss}}|=1/4$ for $i\neq j$.
Accordingly, the most stringent robustness criterion is the survival of a nonzero
long-distance plateau: whether $|C_{m}(r\!\to\!N/2)|_{\mathrm{ss}}$ remains close to
$1/4$ as $r$ approaches the maximal separation.

\subsection{Disorder with negligible impact on the $\protect\eta$-paired NESS%
}

\label{subsec:disorder_robust}

We first discuss disorder types that, as protected by symmetry and confirmed
numerically, leave the steady-state diagnostics essentially invariant (up to
finite-size corrections), merely renormalizing transient relaxation rates.
The defining feature of these perturbations is that they neither open direct
leakage channels out of $\mathcal{H}_{\mathrm{HD}}$ nor generate coherent
torques capable of overcoming the dissipative pinning scale set by the
Liouvillian gap.

\paragraph*{(i) Interaction disorder.}

\begin{figure}[t]
\centering
\includegraphics[width=\linewidth]{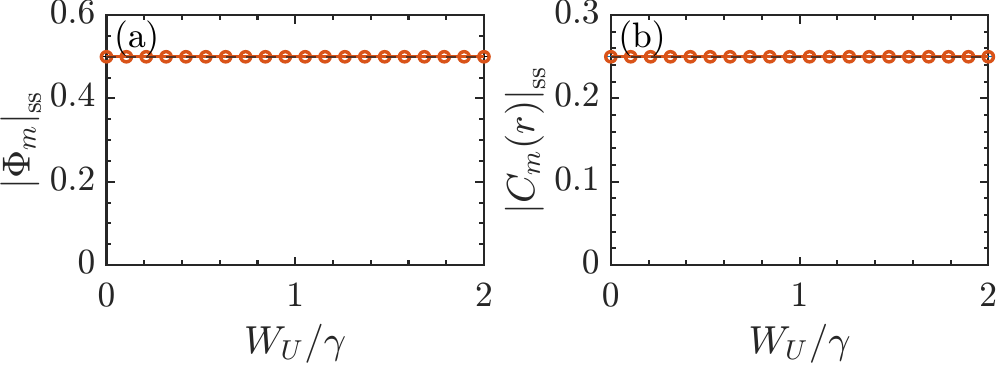}
\caption{ Robustness against interaction disorder $U_i=U+\protect\delta U_i$
as a function of $W_U/\protect\gamma$. (a) $|\Phi_m|_{\mathrm{ss}}$ and (b) $%
|C_m(r\!\to\!N/2)|_{\mathrm{ss}}$. Points show disorder averages over $100$
realizations; dashed lines mark the projected benchmarks $1/2$ and $1/4$.
Parameters: 1D chain with PBC, $t=\protect\gamma$, $U=8\gamma$, $%
\protect\theta=\protect\pi/2$, $N=5$, and dissipation applied to site $1$.
Both correlators remain close to their projected values over the explored
range of $W_U/\protect\gamma$, indicating that the $\protect\eta$-paired
NESS is weakly affected by interaction inhomogeneity. }
\label{fig:interaction_disorder_PRB}
\end{figure}

We first consider randomness in the on-site interaction, 
\begin{eqnarray}
H_{U} &=&\sum_{i}\delta U_{i}\left( n_{i\uparrow }-\frac{1}{2}\right) \left(
n_{i\downarrow }-\frac{1}{2}\right) ,\qquad  \label{eq:disorder_interaction}
\\
\delta U_{i} &\in &[U-W_{U},\,U+W_{U}].
\end{eqnarray}%
This perturbation is diagonal in the local Fock basis and therefore does
not, by itself, create matrix elements between the holon--doublon doublet $%
\{|0\rangle _{i},|\uparrow \downarrow \rangle _{i}\}$ and the singlon states 
$\{|\uparrow \rangle _{i},|\downarrow \rangle _{i}\}$. More importantly for
our protocol, in the particle--hole symmetric convention the interaction
term acts trivially within the holon--doublon manifold: on both $|0\rangle
_{i}$ and $|\uparrow \downarrow \rangle _{i}$ one has 
\begin{equation}
\left( n_{i\uparrow }-\frac{1}{2}\right) \left( n_{i\downarrow }-\frac{1}{2}%
\right) = \frac{1}{4},  \label{eq:U_eigs_PHS}
\end{equation}%
whereas on singlons it equals $-\frac{1}{4}$. Therefore, projecting onto the
HD sector yields only an additive constant, 
\begin{equation}
P_{\mathrm{HD}}\,H_{U}\,P_{\mathrm{HD}}=\sum_{i}\delta U_{i}\,\frac{1}{4}%
\,P_{\mathrm{HD}}=\mathrm{const.}\times P_{\mathrm{HD}},
\label{eq:HU_project_HD_constant}
\end{equation}%
i.e.\ interaction inhomogeneity does not generate an effective longitudinal $%
\eta ^{z}$ field nor any coherent precession within $\mathcal{H}_{\mathrm{HD}%
}$ (and hence within the projected doubled manifold).

As a result, $U$-disorder does not compete with the dissipative axis
selection: it neither tilts the locally stabilized $+y$ dark direction nor
enhances leakage out of the reduced manifold at leading order. The steady
state selected by the rotated jumps is therefore unchanged. Consistent with
this microscopic observation, our numerics show essentially no dependence of
the NESS plateau values on the disorder strength $W_U$ in the explored
window: both $|\Phi_m|_{\mathrm{ss}}$ and $|C_m(r\!\to\!N/2)|_{\mathrm{ss}}$
remain pinned to the projected benchmarks $1/2$ and $1/4$ (up to finite-size
fluctuations); see Fig.~\ref{fig:interaction_disorder_PRB}.

\paragraph*{(ii) Inhomogeneous dissipative strength.}

\begin{figure}[t]
\centering
\includegraphics[width=\linewidth]{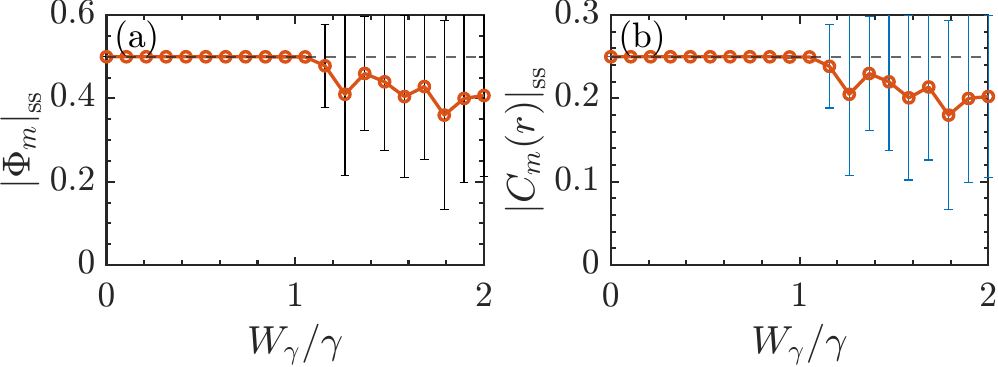}
\caption{ Robustness against dissipation-strength disorder $\protect\gamma_i=%
\protect\gamma+\protect\delta\protect\gamma_i$ as a function of $W_{\protect%
\gamma}/\protect\gamma$. (a) $|\Phi_m|_{\mathrm{ss}}$, and (b) $%
|C_m(r\!\to\!N/2)|_{\mathrm{ss}}$. Points show disorder averages over $100$
realizations; dashed lines mark the projected benchmarks $1/2$ and $1/4$.
Parameters: 1D chain with PBC, $t=\protect\gamma$, $U=8\gamma$, $%
\protect\theta=\protect\pi/2$, $N=5$, and dissipation applied to site $1$.
Both correlators remain close to their projected values over a broad range
of $W_{\protect\gamma}/\protect\gamma$, indicating that dissipative
inhomogeneity weakly renormalizes the projected generator $\mathcal{L}_{%
\mathrm{eff}}^{(0)}$. Error bars denote the standard error of the mean. }
\label{fig:gamma_disorder_PRB}
\end{figure}

We next consider inhomogeneous dissipation strengths, 
\begin{equation}
L_{j}=\sqrt{\gamma _{j}}\,\eta _{j}^{-}e^{i\frac{\pi }{2}\eta
_{j}^{x}},\qquad \gamma _{j}\in \lbrack \gamma -W_{\gamma },\,\gamma
+W_{\gamma }],  \label{eq:disorder_gamma}
\end{equation}%
with the same rotated structure as in the clean protocol. Importantly, the
local dark direction is independent of the pumping rate: for any $\gamma
_{j}>0$ the $+y$-polarized pseudospin is annihilated, $L_{j}\,|\eta
_{j}^{y}=+1/2\rangle =0$. Therefore, $\gamma $-disorder does not shift the
location of the attractive fixed point; it only redistributes the decay
rates that govern the approach to it.

This redistribution has a transparent spectral consequence. In the projected
picture each driven site contributes a local dissipative gap of order $%
\gamma _{j}/2$, so the slowest relaxation channel is controlled by the
weakest pump, $\Delta _{\min }\sim \frac{1}{2}\min_{j\in \mathcal{J}}\gamma
_{j}$. Hence, for moderate disorder where all $\gamma _{j}$ remain
appreciable, the NESS plateau values are essentially unchanged and only the
transient dynamics becomes multi-rate (different sites lock to the $+y$ axis
at different speeds). By contrast, once the distribution becomes broad
enough that some $\gamma _{j}$ are parametrically small (or approach zero
when $W_{\gamma }\gtrsim \gamma $), the effective pinning gap no longer
provides a uniform lower bound. In that regime, otherwise subleading
processes, such as hopping-induced virtual excursions outside the reduced
manifold, can operate over the long times set by $1/\Delta _{\min }$ and
lead to visible deviations from the ideal $\eta $-paired plateau in finite
systems. Numerically, this crossover occurs when $W_{\gamma }$ becomes
comparable to $\gamma $; see Fig.~\ref{fig:gamma_disorder_PRB}.

\paragraph*{(iii) Spin-dependent (magnetic) disorder (Zeeman fields).}

\begin{figure}[t]
\centering
\includegraphics[width=\linewidth]{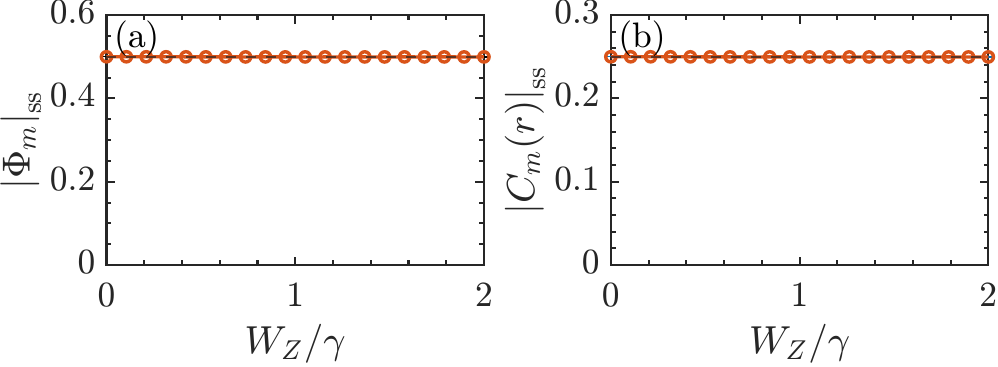}
\caption{ Robustness against Zeeman-field disorder $B_{i}^{z}%
\,S_{i}^{z}+B_{i}^{x}\,S_{i}^{x}$ as a function of $W_{Z}/\protect\gamma$.
(a1) $|\Phi_m|_{\mathrm{ss}}$, and (b1) $|C_m(r\!\to\!N/2)|_{\mathrm{ss}}$.
Points show disorder averages over $100$ realizations; dashed lines mark the
projected benchmarks $1/2$ and $1/4$. Parameters: 1D chain with PBC, $t=%
\protect\gamma$, $U=8\gamma$, $\protect\theta=\protect\pi/2$, $N=5$%
, and dissipation applied to site $1$. Both correlators remain close to the
projected values over the explored range of $W_Z/\protect\gamma$, indicating
that the $\protect\eta$-paired NESS is only weakly affected by Zeeman
inhomogeneity. }
\label{fig:zeeman_disorder_PRB}
\end{figure}
We now emphasize a point that is crucial for interpreting the numerics: a
random Zeeman field, 
\begin{eqnarray}
H_{Z} &=&\sum_{i}\left( B_{i}^{z}\,S_{i}^{z}+B_{i}^{x}\,S_{i}^{x}\right) , \\
B_{i}^{z},\text{ }B_{i}^{x} &\in &[-W_{Z},\,W_{Z}],
\end{eqnarray}%
does not compete with $\eta $-pair pumping in our protocol. The reason is
structural: the Hubbard model has $\mathrm{SU}(2)_{S}\times \mathrm{SU}%
(2)_{\eta }$ symmetry, and the two algebras commute, $[S_{i}^{\alpha },\eta
_{j}^{\beta }]=0$ for all $i,j$. Consequently, 
\begin{equation}
\lbrack H_{Z},\eta _{i}^{\pm }]=[H_{Z},\eta _{i}^{x,y,z}]=0,
\end{equation}%
and, even more directly, $H_{Z}$ acts trivially on the local HD doublet $%
\{|0\rangle _{i},|\uparrow \downarrow \rangle _{i}\}$ because both states
are spin singlets ($\vec{S}_{i}=0$). Therefore Zeeman disorder neither
breaks the HD closure nor introduces an axis competing with the dissipative $%
+y$ pinning. In the full microscopic dynamics, $H_{Z}$ can only enter
through higher-order virtual processes together with hopping, which
renormalize transients but do not move the attractive fixed point selected
by the dissipator. Numerically, we indeed find that the two steady-state
diagnostics remain pinned, with $|C_{m}(r\rightarrow N/2)|_{\mathrm{ss}%
}\rightarrow 1/4$ and $|\Phi _{m}|_{\mathrm{ss}}\rightarrow 1/2$ across the
explored field strengths; see Fig.~\ref{fig:zeeman_disorder_PRB}.

\paragraph*{(iv) Moderate bond disorder in the Zeno/large-$U$ window.}

\begin{figure}[t]
\centering
\includegraphics[width=\linewidth]{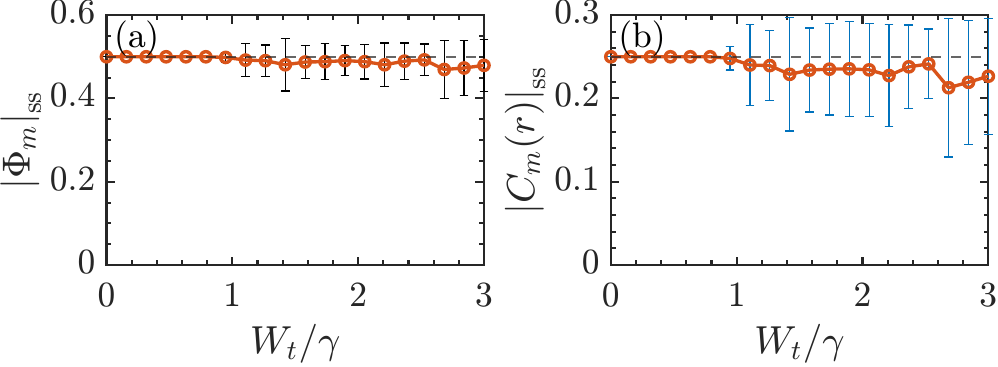}
\caption{ Robustness against bond disorder $\protect\delta t_{ij}$ as a function of
$W_t/\protect\gamma$. (a) $|\Phi_m|_{\mathrm{ss}}$ and (b)
$|C_m(r\!\to\!N/2)|_{\mathrm{ss}}$. Points show disorder averages over 100
realizations, and dashed lines mark the projected benchmarks $1/2$ and $1/4$.
Parameters are the same as in Fig.~\protect\ref{fig:zeeman_disorder_PRB}, but
with no Zeeman field. In the perturbative window set by the large-$U$
separation and/or Zeno pumping, bond-disorder-induced excursions into the
singlon sector remain short-lived and can be treated as virtual processes in
a controlled Schur-complement expansion. Their main effect is therefore to
renormalize relaxation pathways within the reduced holon--doublon manifold.
Accordingly, both correlators stay close to the projected values over a wide
range of $W_t/\protect\gamma$. For stronger bond disorder, sample-to-sample
fluctuations grow and the long-distance plateau is gradually suppressed,
signaling the buildup of real singlon population and the breakdown of the
projected description.
}
\label{fig:bond_disorder_PRB}
\end{figure}
We finally consider randomness in the kinetic energy, 
\begin{equation}
H_{t,\mathrm{dis}}=-\sum_{\langle ij\rangle ,\sigma }\delta t_{ij}\bigl(%
c_{i\sigma }^{\dagger }c_{j\sigma }+\mathrm{H.c.}\bigr),\qquad \delta
t_{ij}\in [t-W_{t},\,t+W_{t}],  \label{eq:disorder_bond}
\end{equation}
which is the most direct microscopic perturbation that promotes processes
passing through singly occupied (singlon) configurations. Indeed, a hop
starting from a holon--doublon configuration necessarily generates
intermediate single occupancies on the two sites involved, so $H_{t,\mathrm{%
dis}}$ does not preserve the holon--doublon (HD) sector as a strict
invariant subspace.

A crucial subtlety is that, for the particle--hole symmetric Hubbard model on a
bipartite lattice, inhomogeneous A--B hopping, as long as it does not introduce
frustrating same-sublattice terms, still does not generate coherent dynamics
within the $\eta$-multiplet. In the reduced description this appears as a
vanishing projected Hamiltonian contribution,
$P[-iH_{\mathrm{Hub}},\cdot]P=0$ on $\mathcal{H}_{\Psi}\otimes\mathcal{H}_{\Psi}^{\ast}$,
so the leading projected generator $\mathcal{L}_{\mathrm{eff}}^{(0)}$ keeps the
same dark-state structure (Appendix~\ref{app:eff_L_derivation}). As a result,
bond disorder does not tilt the dissipative axis selected inside the reduced
manifold, in contrast to angle disorder.

Bond disorder mainly influences the dynamics through the off-manifold couplings
that connect the target subspace to singlon sectors, i.e., the $P\mathcal{L}Q$
and $Q\mathcal{L}P$ blocks in the Schur-complement elimination. A broader spread
of $t_{ij}$ strengthens and spatially inhomogenizes these couplings, which in
turn enhances higher-order corrections such as
$-P\mathcal{L}Q\,(Q\mathcal{L}Q)^{-1}Q\mathcal{L}P+\cdots$. In the regime where
these excursions remain perturbative and short-lived (here controlled by the
dissipative separation induced by the driven sites), the main effect is to
reroute relaxation and broaden transient time scales, while leaving the steady
state essentially unchanged. Accordingly, both $|\Phi_m|_{\mathrm{ss}}$ and the
long-distance plateau $|C_m(r\!\to\!N/2)|_{\mathrm{ss}}$ stay close to the
projected benchmarks $1/2$ and $1/4$ over a wide range of $W_t$ (up to
finite-size effects); see Fig.~\ref{fig:bond_disorder_PRB}.

This robustness has a clear limit. Once bond disorder is strong enough that
hopping-induced excursions are no longer perturbative on the dissipative
separation scale, singlon population can build up and the projected description
ceases to apply. Numerically, this crossover appears as a visible suppression
of the ODLRO plateau $|C_m(r\!\to\!N/2)|_{\mathrm{ss}}$---typically stronger than
the reduction of the one-point amplitude $|\Phi_m|_{\mathrm{ss}}$---together
with larger sample-to-sample fluctuations at large $W_t$, as shown in
Fig.~\ref{fig:bond_disorder_PRB}.

\subsection{Disorder that suppresses or destroys the $\protect\eta$-pair
ODLRO plateau}

\label{subsec:disorder_destructive}

We now turn to disorders that do affect the steady-state correlators.
Operationally, these perturbations either (i) generate a coherent torque
within $\mathcal{H}_{\mathrm{HD}}$ that competes with the dissipative
pinning axis, reducing $|\Phi _{m}|_{\mathrm{ss}}$ and $|C_{m}(r\rightarrow
N/2)|_{\mathrm{ss}}$, or (ii) open genuinely nonperturbative leakage
pathways (e.g. resonant singlon participation or pair-breaking channels)
that invalidate the effective pumping picture and drive $|C_{m}(r\rightarrow
N/2)|_{\mathrm{ss}}\rightarrow 0$.

\paragraph*{(i) On-site potential disorder at half filling.}

\begin{figure}[t]
\centering
\includegraphics[width=\linewidth]{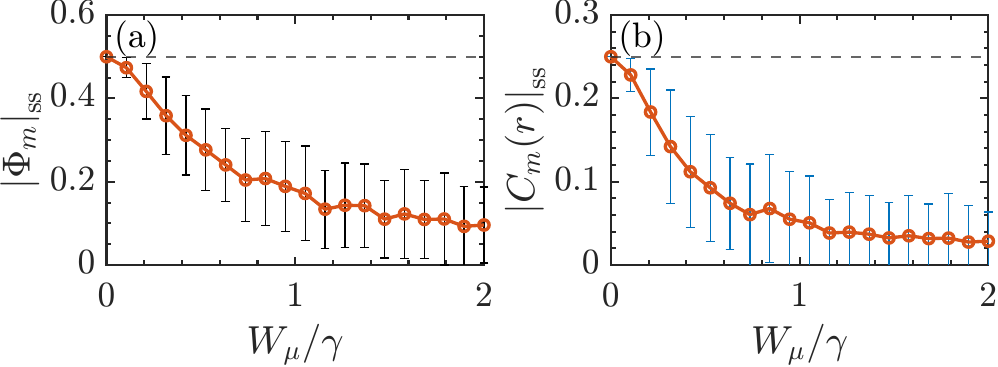}
\caption{ Response to on-site (chemical-potential) disorder $\protect\mu_i=%
\protect\mu+\protect\delta\protect\mu_i$ as a function of $W_{\protect\mu}/%
\protect\gamma$. (a) $|\Phi_m|_{\mathrm{ss}}$, and (b) $|C_m(r\!\to\!N/2)|_{%
\mathrm{ss}}$. Parameters: 1D chain with PBC, $t=\protect\gamma$, $U=6\gamma$, $\protect\theta=\protect\pi/2$, $N=5$, and dissipation
applied on site $1$. Points show disorder averages over $100$ realizations;
dashed lines mark the projected benchmarks $1/2$ and $1/4$. In contrast to
symmetry-compatible perturbations, on-site inhomogeneity acts as an
effective longitudinal field in the $\protect\eta$ sector and competes with
the dissipative phase locking along $+y$. Consistently, both correlators
decrease rapidly with increasing $W_{\protect\mu}/\protect\gamma$, and the
long-distance plateau is strongly suppressed at moderate disorder,
indicating the breakdown of the projected-manifold picture and the loss of
robust ODLRO.}
\label{fig:mu_disorder_PRB}
\end{figure}

A random scalar potential, 
\begin{eqnarray}
H_{\mu } &=&\sum_{i}\mu _{i}\,(n_{i\uparrow }+n_{i\downarrow }-1)
\label{eq:disorder_mu} \\
&=&2\sum_{i}\mu _{i}\,\eta _{i}^{z}, \\
\mu _{i} &\in &[-W_{\mu },\,W_{\mu }],
\end{eqnarray}%
is qualitatively more dangerous than the previous perturbations because it
does not merely broaden decay rates or renormalize phases inside the HD
sector: once combined with hopping, it can activate efficient pathways that
populate singlons and thereby undermines the mechanism behind the $\eta $%
-pair pumping.

At the single-site level, $H_{\mu}$ acts as a random longitudinal field within
the local $\eta$ doublet and detunes the holon and doublon components through
$\eta_i^{z}$. This term is diagonal in the local Fock basis, so by itself it
cannot populate the singlon sector. Its impact arises in the full Hubbard
model, where hopping $H_t$ couples the holon--doublon manifold to intermediate
singly occupied states via virtual processes. The corresponding energy
denominators are set by local many-body splittings, which are reshuffled by
the site-dependent potential offsets. As the disorder strength $W_{\mu}$
increases, near-resonant configurations become more frequent and the
singlon-assisted processes are no longer suppressed by large denominators.
As a result, $H_{\mu}$ does not create singlons directly, but it strongly
enhances the singlon-mediated mixing generated by $H_t$ by weakening, and in
rare regions effectively removing, the protection provided by the interaction
gap and by the dissipative separation.

From the viewpoint of the effective-generator expansion based on Schur-complement
elimination, this is the regime where the off-manifold resolvent becomes large.
The operator $(Q\hat{\mathcal{L}}Q)^{-1}$ is no longer parametrically small,
because $H_{\mu}$ can shift parts of the $Q$-sector spectrum toward zero and
thereby promote terms that were higher order in the expansion to leading
contributions. Once this happens, the rotated dissipation can no longer be viewed as an
approximately closed pumping problem within the holon--doublon manifold or the
$\eta$-paired sector. The attractive fixed point is then destabilized by real
leakage into singlon configurations.

Numerically, this crossover shows up as a clear suppression of both the
steady-state one-point amplitude $|\Phi_{m}|_{\mathrm{ss}}$ and the
long-distance plateau $|C_{m}(r\!\to\!N/2)|_{\mathrm{ss}}$ as $W_{\mu}$ is
increased. In this regime the ODLRO plateau collapses, as illustrated in
Fig.~\ref{fig:mu_disorder_PRB}.

\paragraph*{(ii) Transverse pseudospin fields within $\mathcal{H}_{\mathrm{HD%
}}$.}

\begin{figure}[t]
\centering
\includegraphics[width=\linewidth]{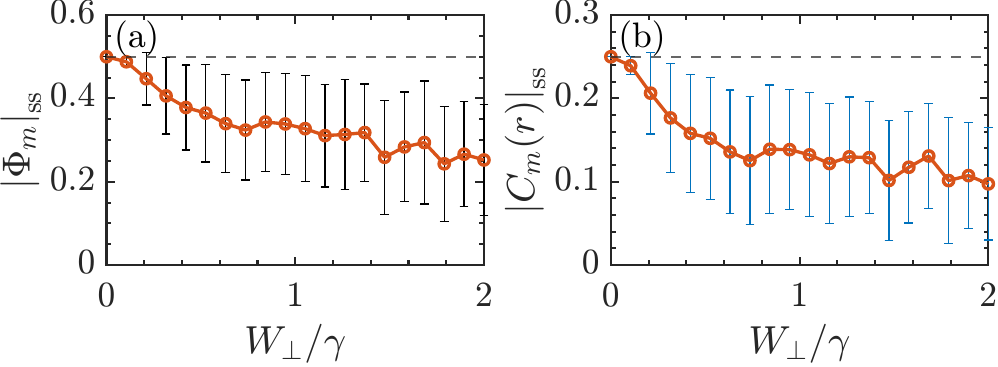}
\caption{ Response to a random transverse field acting directly in the
holon--doublon manifold, $H_\perp=\sum_i(h_i^x\,\protect\eta_i^x+h_i^y\,%
\protect\eta_i^y)$ with $h_i^{x,y}\in[-W_\perp,W_\perp]$, as a function of $%
W_\perp/\protect\gamma$. (a) $|\Phi_m|_{\mathrm{ss}}$, and (b) $%
|C_m(r\!\to\!N/2)|_{\mathrm{ss}}$. Parameters: 1D chain with PBC, $t=\protect%
\gamma$, $U=6\gamma$, $\protect\theta=\protect\pi/2$, $N=5$, and
dissipation applied on site $1$. Points show disorder averages over $100$
realizations; dashed lines mark the projected benchmarks $1/2$ and $1/4$.
Unlike potential disorder, $H_\perp$ is already operative within $\mathcal{H}%
_{\mathrm{HD}}$: it coherently rotates the local $\protect\eta$ pseudospin
away from the dissipatively selected $+y$ axis. Consistently, both
correlators are strongly reduced with increasing $W_\perp/\protect\gamma$,
demonstrating direct competition with the dissipative phase locking and a
rapid suppression of the ODLRO plateau.}
\label{fig:perp_disorder_PRB}
\end{figure}
A random transverse field acting directly within the holon--doublon manifold, 
\begin{eqnarray}
H_{\perp} &=& \sum_i \bigl(h_i^x\,\eta_i^x+h_i^y\,\eta_i^y\bigr), \label{eq:disorder_transverse_eta} \\
h_i^{x,y} &\in& [-W_{\perp}, W_{\perp}],
\end{eqnarray}
represents a qualitatively different and more dangerous perturbation than potential disorder. While potential disorder becomes destructive primarily by enhancing singlon-mediated leakage through hopping, $H_{\perp}$ operates directly within $\mathcal{H}_{\mathrm{HD}}$ by coherently rotating the local $\eta$-pseudospin away from the dissipatively selected $+y$ axis.

Numerically, this competition is immediately evident. As soon as $H_{\perp} \neq 0$, both the steady-state one-point amplitude $|\Phi_m|_{\mathrm{ss}}$ and the long-distance plateau $|C_m(r \to N/2)|_{\mathrm{ss}}$ depart from the projected benchmarks, decreasing monotonically with increasing $W_{\perp}$, as shown in Fig.~\ref{fig:perp_disorder_PRB}. Notably, the ODLRO plateau $C_{\infty} \simeq 1/4$ characteristic of the ideal projected theory is compromised even at small $W_{\perp}$, indicating that the $\eta$-paired NESS is not perturbatively stable against transverse fields.

The microscopic origin of this sensitivity is straightforward. The clean protocol relies on dissipative axis selection, where local jumps pin each site toward the $+y$ dark state. A transverse Hamiltonian field continuously generates coherent precession within $\mathcal{H}_{\mathrm{HD}}$, which prevents the dark condition from being dynamically preserved and forces the steady state away from the $+y$ product fixed point. The resulting NESS becomes a mixed state determined by the competition between coherent rotation and dissipative repumping, leading to a substantial reduction in long-distance $\eta$-coherence.

\paragraph*{(iii) Rotation-angle disorder in the jump operators.}

\begin{figure}[t]
\centering
\includegraphics[width=\linewidth]{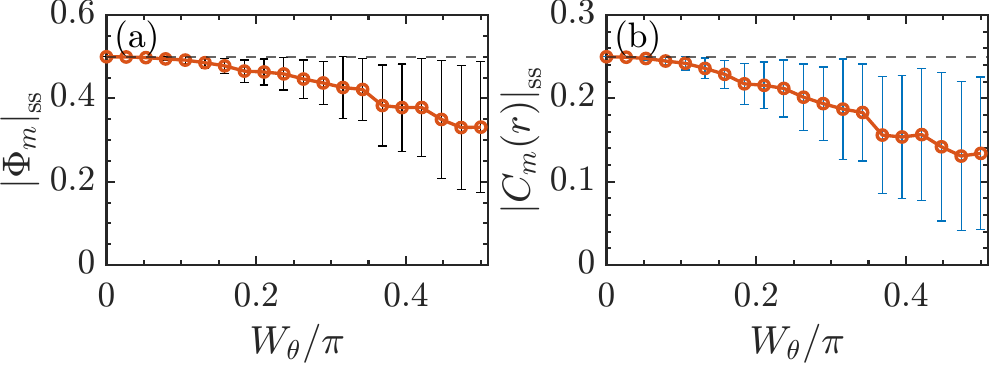}
\caption{ Response to angle disorder in the rotated dissipation, $\protect%
\theta_j=\protect\pi/2+\protect\delta\protect\theta_j$ with $\protect\delta%
\protect\theta_j\in[-W_\protect\theta,W_\protect\theta]$, as a function of $%
W_\protect\theta/\protect\pi$. (a) $|\Phi_m|_{\mathrm{ss}}$ and, (b) $%
|C_m(r\!\to\!N/2)|_{\mathrm{ss}}$. Parameters: 1D chain with PBC, $t=\protect%
\gamma$, $U=6\gamma$, nominal rotation $\protect\theta=\protect\pi/2
$, $N=5$, and dissipation applied on site $1$. Points show disorder averages
over $100$ realizations; dashed lines mark the projected benchmarks $1/2$
and $1/4$. Angle disorder tilts the locally selected dissipative axis: the
clean $+y$ dark state is no longer exactly annihilated for $\protect\delta%
\protect\theta_j\neq0$, so the attractive fixed point is shifted. Consistent
with this, the steady-state correlators remain close to the projected values
at weak disorder (small $W_\protect\theta/\protect\pi$), but are
progressively suppressed as $W_\protect\theta/\protect\pi$ increases, with
the long-distance correlator $|C_m|_{\mathrm{ss}}$ being more sensitive than
the one-point amplitude $|\Phi_m|_{\mathrm{ss}}$.}
\label{fig:theta_disorder_PRB}
\end{figure}
Imperfect feedback rotation, described by the modified jump operators
\begin{eqnarray}
L_{j} &=&\sqrt{\gamma }\,\eta _{j}^{-}e^{i\theta _{j}\eta _{j}^{x}}, \\
\theta _{j} &=&\frac{\pi }{2}+\delta \theta _{j}, \quad \delta \theta _{j} \in [-W_{\theta }, W_{\theta }],
\end{eqnarray}
represents a direct perturbation of the pumping mechanism itself. This disorder rotates the jump operator away from the ideal $\pi /2$ frame, thereby shifting the locally selected dissipative axis. Consequently, the ideal dark state $|\eta_{j}^{y}=+1/2\rangle$ is no longer annihilated when $\delta \theta _{j} \neq 0$. 

Even weak angle disorder immediately displaces the attractive fixed point, causing the NESS to depart from the pure $+y$ product state. In numerical simulations, this mechanism manifests as a rapid decrease in both the steady-state one-point amplitude $|\Phi _{m}|_{\mathrm{ss}}$ and the long-distance plateau $|C_{m}(r \rightarrow N/2)|_{\mathrm{ss}}$ as $W_{\theta }$ increases, as illustrated in Fig.~\ref{fig:theta_disorder_PRB}.

\paragraph*{(iv) Pair-breaking noise outside the designed channel.}

\begin{figure}[t]
\centering
\includegraphics[width=\linewidth]{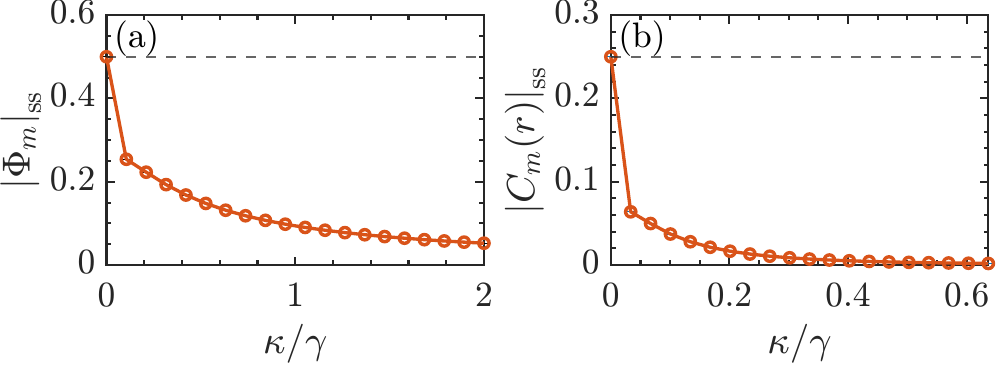}
\caption{ Pair-breaking single-particle loss with rate $\protect\kappa$
(normalized as $\protect\kappa/\protect\gamma$) strongly suppresses the $%
\protect\eta$-paired NESS. (a) $|\Phi_m|_{\mathrm{ss}}$ and (b) the
long-distance plateau $|C_m(r\!\to\!N/2)|_{\mathrm{ss}}$ versus $\protect%
\kappa/\protect\gamma$. Dashed lines indicate the projected benchmarks $1/2$
and $1/4$ at $\protect\kappa=0$. Parameters: 1D chain with PBC, $t=\protect%
\gamma$, $U=6\gamma$, $\protect\theta=\protect\pi/2$, $N=5$, and
dissipation applied on site $2$. Turning on $\protect\kappa$ induces a
leak-and-scramble mechanism: $|C_m(r\!\to\!N/2)|_{\mathrm{ss}}$ collapses
rapidly toward zero, while $|\Phi_m|_{\mathrm{ss}}$ is reduced more
gradually due to short-time local repumping but remains bounded by the
continuous pair breaking. }
\label{fig:kappa_loss_PRB}
\end{figure}

Finally, we consider additional uncontrolled dissipators that are not aligned with the $\eta$-pumping protocol, the most prominent example being single-particle loss, 
\begin{eqnarray}
L_{i\sigma }^{\mathrm{loss}} &=&\sqrt{\kappa }\,c_{i\sigma }, \\
L_{i\sigma }^{\mathrm{dep}} &=&\sqrt{\kappa_{\phi }}\,n_{i\sigma },
\end{eqnarray}
where the latter represents local dephasing. These channels are qualitatively distinct from the disorder terms previously discussed because they act as genuine pair-breaking processes. In the local Fock basis, $c_{i\sigma}$ converts holons and doublons into singly occupied states,
\begin{equation}
c_{i\sigma }: |0\rangle _{i} \mapsto 0, \quad |\uparrow \downarrow \rangle _{i} \mapsto |\bar{\sigma}\rangle _{i}, \qquad c_{i\sigma }^{\dagger }: |0\rangle _{i} \mapsto |\sigma \rangle _{i}.
\end{equation}
Such processes open real dissipative leakage channels from the holon--doublon doublet into the singlon sector. In the $\eta$-pseudospin language, this corresponds to the irreversible destruction of local coherence. While the designed jump $L_{j}$ can rephase and repump within the $\eta _{i}=1/2$ doublet, single-particle loss continuously depletes the weight of that sector and scrambles the relative phase between $|0\rangle _{i}$ and $|\uparrow \downarrow \rangle _{i}$.

Numerically, this leak-and-scramble mechanism is highly detrimental to $\eta$-pair coherence. As soon as the loss rate $\kappa$ becomes nonzero, the long-distance correlator $|C_m(r \to N/2)|_{\mathrm{ss}}$ is rapidly suppressed, signaling the prompt destruction of ODLRO as shown in Fig.~\ref{fig:kappa_loss_PRB}. The one-point amplitude $|\Phi_m|_{\mathrm{ss}}$ is also reduced, albeit typically more gradually, as the local rotated pumping can transiently repolarize the $\eta$-pseudospin and maintain short-range coherence. However, the pumping cannot compensate for an ongoing pair-breaking process at long times. Consequently, uncontrolled loss imposes a stringent upper bound on the steady-state ODLRO plateau and, for a fixed pumping strength $\gamma$, determines the achievable quality and effective lifetime of the $\eta$-paired NESS.

In summary, the response to disorder can be organized according to a symmetry-based robustness principle characterized by two complementary failure modes. Perturbations that act diagonally, or effectively longitudinally, within the local holon--doublon doublet typically do not compete with the dissipative axis selection. As long as off-manifold excursions remain perturbative, these terms primarily renormalize relaxation pathways while leaving the ODLRO plateau near its ideal projected value. This category includes bond inhomogeneity of bipartite $A$--$B$ hopping, which fails to generate coherent dynamics within the $\eta$-multiplet at leading order. By contrast, the nonequilibrium superconducting order is suppressed and eventually destroyed when perturbations either tilt the local dissipative axis within the reduced manifold, such as in the case of transverse fields and angle disorder, or cause the off-manifold resolvent to become large. The latter case opens real leakage and pair-breaking channels, as observed with strong potential disorder and single-particle loss.

\section{Conclusion and Outlook}

\label{sec: conclusion}

In this work, we have demonstrated that superconducting-like order can be
stabilized as a nonequilibrium attractor in a strongly correlated lattice
system. For the particle--hole symmetric Hubbard chain, we introduced a
rotated quantum-jump protocol based on a locally rotated $\eta$-pair lowering
operator. We showed that the resulting Lindblad dynamics pumps the system
from the vacuum into a nonequilibrium steady state (NESS) characterized by
$\eta$-pair off-diagonal long-range order (ODLRO).

The mechanism can be summarized as follows. The rotated jump reshapes the
local dark-state structure and pins a definite $\eta$-pseudospin phase, thereby
selecting coherent holon--doublon superpositions as local building blocks. In
the strong-coupling regime, singlon configurations are not populated by the
dissipator to leading order and enter only through hopping-induced virtual
excursions, which remain perturbative and can be eliminated using a controlled
Schur-complement expansion. Coherent hopping then propagates the imprinted
pseudospin phase across the lattice, while the projected Liouvillian dynamics
selects the $\eta$-paired manifold as an attractive sector and opens a
dissipative gap that sets the relaxation timescale. A notable consequence is a
local-to-global preparation principle: dissipation applied on a single site
can induce macroscopic ODLRO throughout the interacting system.

We further found that the $\eta$-paired NESS is robust against broad classes of
static disorder, which mainly renormalize transients and relaxation rates, and
we identified perturbations that suppress long-range order by competing with
dissipative axis selection or by opening genuine leakage and pair-breaking
channels.

Looking forward, it will be valuable to develop analytic scaling theories for
the dissipative gap and relaxation as functions of system size, driving
geometry, and disorder strength. It is also natural to explore extensions to
higher-dimensional bipartite lattices and to investigate spatially patterned
or time-dependent rotations and jumps as tools to imprint phase textures,
domains, and currents in a controlled and autonomous manner. More broadly, our results highlight minimal local dissipation as a practical route to stabilizing correlated order in driven open systems.

\acknowledgments We acknowledge the support of the National Natural Science
Foundation of China under Grants No. 12275193 and 11975166.

\appendix

\section{Explicit Liouvillian matrix for the single-qubit example}

\label{app:single_qubit_liouvillian}

In this Appendix we present the explicit matrix representation of the
Liouvillian for the single-qubit models in Sec.~\ref%
{sec:single_spin_heuristic}. We employ the standard vectorization identity $%
\mathrm{vec}(A\rho B)=(A\otimes B^{T})\mathrm{vec}(\rho )$, under which 
\begin{equation}
\mathcal{L}=(L\otimes L^{\ast })-\frac{1}{2}(L^{\dagger }L\otimes \mathbb{I)}%
-\frac{1}{2}(\mathbb{I}\otimes (L^{\dagger }L)^{T}),  \label{eq:vec_L_H0}
\end{equation}%
where $H=0$ has been used.

We work in the $z$-basis $\{|\uparrow\rangle,|\downarrow\rangle\}$ with 
\begin{equation}
s^{x}=\frac{1}{2} 
\begin{pmatrix}
0 & 1 \\ 
1 & 0%
\end{pmatrix}%
, \qquad s^{y}=\frac{1}{2} 
\begin{pmatrix}
0 & -i \\ 
i & 0%
\end{pmatrix}%
, \qquad s^{-}= 
\begin{pmatrix}
0 & 0 \\ 
1 & 0%
\end{pmatrix}%
.
\end{equation}
We choose the ordering $\mathrm{vec}(\rho)=(\rho_{\uparrow\uparrow},\rho_{%
\uparrow\downarrow},\rho_{\downarrow\uparrow},\rho_{\downarrow%
\downarrow})^{T}$.

For the rotated jump $L(\theta )=\sqrt{\gamma }\,s^{-}e^{i\theta s^{x}}$,
the case $\theta =\pi /2$ yields 
\begin{equation}
s^{-}e^{i\frac{\pi }{2}s^{x}}=\frac{\sqrt{2}}{2}%
\begin{pmatrix}
0 & 0 \\ 
1 & i%
\end{pmatrix}%
,
\end{equation}%
and Eq.~\ref{eq:vec_L_H0} gives the explicit $4\times 4$ Liouvillian matrix 
\begin{equation}
\mathcal{L}_{\pi /2}=\gamma 
\begin{pmatrix}
-\frac{1}{2} & \frac{i}{4} & -\frac{i}{4} & 0 \\ 
-\frac{i}{4} & -\frac{1}{2} & 0 & -\frac{i}{4} \\ 
\frac{i}{4} & 0 & -\frac{1}{2} & \frac{i}{4} \\ 
\frac{1}{2} & -\frac{i}{4} & \frac{i}{4} & 0%
\end{pmatrix}%
.  \label{eq:Lpi2_matrix}
\end{equation}%
Its spectrum contains a unique mode with $\mathrm{Re}(\lambda )=0$ (indeed $%
\lambda =0$), which determines the steady state. Solving $\mathcal{L}_{\pi
/2}\,\mathrm{vec}(\rho _{\mathrm{ss}})=0$ yields 
\begin{equation}
\mathrm{vec}(\rho _{\mathrm{ss}})\propto 
\begin{pmatrix}
1 \\ 
-i \\ 
i \\ 
1%
\end{pmatrix}%
,\qquad \rho _{\mathrm{ss}}\propto 
\begin{pmatrix}
1 & -i \\ 
i & 1%
\end{pmatrix}%
=2\,|\psi _{\mathrm{d}}(\tfrac{\pi }{2})\rangle \langle \psi _{\mathrm{d}}(%
\tfrac{\pi }{2})|,
\end{equation}%
which is consistent with the pure dark state in Eq.~\ref%
{eq:dark_state_rotated}.

\section{Derivation of the projected effective Liouvillian from Feshbach
(Schur-complement) elimination}

\label{app:eff_L_derivation}

This appendix derives the projected generator Eq.~%
\eqref{eq:exact_liouvillian_form} starting from the microscopic Lindblad
equation \eqref{eq:lindblad_master}. Our goal is twofold: (i) to obtain the
leading effective Liouvillian acting within the chosen slow manifold, and
(ii) to record a systematic organization of higher-order corrections via the
Feshbach (Schur-complement) map.

\subsection{Feshbach map for a Liouvillian superoperator}

We consider the vectorized form of the Lindblad generator, viewed as a
linear operator $\hat{\mathcal{L}}$ acting on the doubled space $\mathcal{H}%
\otimes \mathcal{H}^{\ast }$. Let $P$ be the projector onto the selected
slow manifold $\mathcal{M}$ (in the main text, the doubled manifold relevant
to the $\eta $-pairing discussion), and $Q=\mathbb{I}-P$ be the
complementary projector. Writing the eigenproblem 
\begin{equation}
\hat{\mathcal{L}}\,|x\rangle \rangle =\lambda \,|x\rangle \rangle
\end{equation}%
in the block form with respect to $P\oplus Q$, one obtains the exact
Feshbach map 
\begin{equation}
\hat{\mathcal{L}}_{\mathrm{eff}}(\lambda )=P\hat{\mathcal{L}}P+P\hat{%
\mathcal{L}}Q\,(\lambda -Q\hat{\mathcal{L}}Q)^{-1}\,Q\hat{\mathcal{L}}P,
\label{eq:feshbach_map_general}
\end{equation}%
whenever $(\lambda -Q\hat{\mathcal{L}}Q)$ is invertible on $Q(\mathcal{H}%
\otimes \mathcal{H}^{\ast })$. For the steady-state problem one sets $%
\lambda =0$, yielding the standard expansion 
\begin{equation}
\hat{\mathcal{L}}_{\mathrm{eff}}=P\hat{\mathcal{L}}P-P\hat{\mathcal{L}}Q\,(Q%
\hat{\mathcal{L}}Q)^{-1}\,Q\hat{\mathcal{L}}P+\cdots ,
\label{eq:feshbach_eff_L}
\end{equation}%
where the ellipsis denotes higher-order contributions (and, if $Q\hat{%
\mathcal{L}}Q$ has a kernel, the inverse is understood as an appropriate
pseudo-inverse). Equation~\eqref{eq:feshbach_eff_L} provides the systematic
meaning of the leading projected generator $P\hat{\mathcal{L}}P$ used in
Sec.~\ref{sec:eta_ketbra_section}: it captures the dominant attractive fixed
point and the primary dissipative gap, while the subleading Schur-complement
term encodes perturbative excursions out of $\mathcal{M}$.

\subsection{Vectorization and ket$\otimes$bra representation}

Under the Choi--Jamio\l kowski isomorphism, the density matrix is mapped to
a vector $|\rho \rangle \rangle \in \mathcal{H}\otimes \mathcal{H}^{\ast }$
such that left (right) multiplication becomes an operator acting on the ket
(bra) factor. The microscopic Lindbladian \eqref{eq:lindblad_master} is
represented by the non-Hermitian operator 
\begin{align}
\hat{\mathcal{L}}=& -i\left( H_{\mathrm{Hub}}\otimes \mathbb{I}-\mathbb{I}%
\otimes H_{\mathrm{Hub}}^{T}\right)  \notag \\
& +\sum_{j\in \mathcal{J}}\left( L_{j}\otimes L_{j}^{\ast }-\frac{1}{2}%
L_{j}^{\dagger }L_{j}\otimes \mathbb{I}-\frac{1}{2}\mathbb{I}\otimes
L_{j}^{T}L_{j}^{\ast }\right) .  \label{eq:vectorized_L_general}
\end{align}%
In the main text we choose $\mathcal{M}$ such that, within $P(\mathcal{H}%
\otimes \mathcal{H}^{\ast })$, the coherent part does not contribute: 
\begin{equation}
P\left[ -i\left( H_{\mathrm{Hub}}\otimes \mathbb{I}-\mathbb{I}\otimes H_{%
\mathrm{Hub}}^{T}\right) \right] P=0,
\label{eq:coherent_part_vanish_projected}
\end{equation}%
which follows from the properties of $H_{\mathrm{Hub}}$ on the chosen
manifold (see Sec.~\ref{sec:eta_ketbra_section}). Hence the leading
projected generator $P\hat{\mathcal{L}}P$ is governed purely by the
dissipators. We take the rotated jump operator 
\begin{equation}
L_{j}=\sqrt{\gamma }\,\eta _{j}^{-}e^{i\frac{\pi }{2}\eta _{j}^{x}}.
\end{equation}%
Within the local holon--doublon doublet (a spin-$1/2$ representation of $%
\eta _{j}$), one has the identity 
\begin{equation}
\eta _{j}^{+}\eta _{j}^{-}=\frac{1}{2}-\eta _{j}^{z}.
\label{eq:eta_id_spinhalf_app}
\end{equation}%
Therefore 
\begin{equation*}
L_{j}^{\dagger }L_{j}=\gamma \,e^{-i\frac{\pi }{2}\eta _{j}^{x}}\,\eta
_{j}^{+}\eta _{j}^{-}\,e^{i\frac{\pi }{2}\eta _{j}^{x}}=\gamma \,e^{-i\frac{%
\pi }{2}\eta _{j}^{x}}\left( \frac{1}{2}-\eta _{j}^{z}\right) e^{i\frac{\pi 
}{2}\eta _{j}^{x}}.
\end{equation*}%
Using the rotation identity 
\begin{equation}
e^{-i\frac{\pi }{2}\eta _{j}^{x}}\eta _{j}^{z}e^{i\frac{\pi }{2}\eta
_{j}^{x}}=-\eta _{j}^{y},  \label{eq:rotation_identity_y_app}
\end{equation}%
we obtain the compact form 
\begin{equation}
L_{j}^{\dagger }L_{j}=\frac{\gamma }{2}-\gamma \,\eta _{j}^{y}.
\label{eq:LdL_y_app}
\end{equation}

\subsection{Leading projected generator $P\hat{\mathcal{L}}P$ and Eq.~ 
\eqref{eq:exact_liouvillian_form}}

Substituting Eq.~\eqref{eq:LdL_y_app} into Eq.~%
\eqref{eq:vectorized_L_general}, and denoting operators acting on the bra
factor by tildes, the anticommutator terms become 
\begin{equation}
-\frac{1}{2}L_{j}^{\dagger }L_{j}\otimes \mathbb{I}-\frac{1}{2}\mathbb{I}%
\otimes L_{j}^{T}L_{j}^{\ast }=-\frac{\gamma }{2}\,\mathbb{I}+\frac{\gamma }{%
2}\left( \eta _{j}^{y}-\tilde{\eta}_{j}^{y}\right) .
\label{eq:anticommutator_ketbra_app}
\end{equation}%
The relative sign between $\eta _{j}^{y}$ and $\tilde{\eta}_{j}^{y}$ follows
from $\eta ^{y\ast }=-\eta ^{y}$ in the standard representation (while $\eta
^{x\ast }=\eta ^{x}$). The recycling term reads 
\begin{equation}
L_{j}\otimes L_{j}^{\ast }=\gamma \left( \eta _{j}^{-}e^{i\frac{\pi }{2}\eta
_{j}^{x}}\right) \left( \tilde{\eta}_{j}^{-}e^{-i\frac{\pi }{2}\tilde{\eta}%
_{j}^{x}}\right) ,  \label{eq:jump_ketbra_app}
\end{equation}%
which couples the ket and bra factors locally. Summing over driven sites $%
j\in \mathcal{J}$ and restricting to the leading projected generator yields 
\begin{eqnarray}
\hat{\mathcal{L}}_{\mathrm{eff}}^{(0)} &\equiv &P\hat{\mathcal{L}}P
\label{eq:Leff0_app} \\
&=&-\frac{\gamma |\mathcal{J}|}{2}+\gamma \sum_{j\in \mathcal{J}}\left[ 
\frac{1}{2}\left( \eta _{j}^{y}-\tilde{\eta}_{j}^{y}\right) +\eta
_{j}^{-}e^{i\frac{\pi }{2}\eta _{j}^{x}}\,\tilde{\eta}_{j}^{-}e^{-i\frac{\pi 
}{2}\tilde{\eta}_{j}^{x}}\right] ,
\end{eqnarray}%
which is Eq.~\eqref{eq:exact_liouvillian_form} in the main text. The next
term in Eq.~\eqref{eq:feshbach_eff_L}, 
\begin{equation}
-\;P\hat{\mathcal{L}}Q\,(Q\hat{\mathcal{L}}Q)^{-1}\,Q\hat{\mathcal{L}}P,
\label{eq:schur_correction_app}
\end{equation}%
organizes higher-order excursions out of $\mathcal{M}$ and back. In our
context these excursions include (i) mixing among different global-$\eta $
sectors within holon--doublon space induced by the locality of $L_{j}$, and
(ii) virtual leakage processes mediated primarily by the hopping $H_{t}$
that temporarily populate singly occupied configurations. Equation~%
\eqref{eq:schur_correction_app} provides the controlled bookkeeping for how
such processes renormalize transient relaxation pathways and shift nonzero
decay modes, while the leading attractive fixed point captured by $P\hat{%
\mathcal{L}}P$ remains the dominant organizing principle in the regime
studied.

\section{Proof of the strictly upper-triangular form of $\hat{\mathcal{L}}$
in the $\protect\eta ^{y}$ basis}

\label{app:triangular_proof}

In this Appendix we prove the statement used in the main text: within the $%
\eta$-pairing manifold, the operator $\hat{\mathcal{L}}$ in Eq.~%
\eqref{eq:exact_liouvillian_form} admits a strictly upper-triangular matrix
representation in the doubled-space basis where $\eta_j^y$ and $\tilde{\eta}%
_j^y$ are diagonal. As a consequence, the spectrum is given exactly by the
diagonal part, leading to Eq.~\eqref{eq:exact_liouvillian_form} and the
universal gap $\Delta=\gamma/2$.

\subsection{Single-site lemma: $K=\protect\eta^- e^{i\frac{\protect\pi}{2}%
\protect\eta^x}$ is upper triangular in the $\protect\eta^y$ basis}

On a single site, the local pseudospin is a spin-$1/2$ representation of
SU(2). Let $\{|\!+\!y\rangle ,|\!-\!y\rangle \}$ denote the eigenstates of $%
\eta ^{y}$, 
\begin{equation}
\eta ^{y}|\pm y\rangle =\pm \frac{1}{2}|\pm y\rangle ,\qquad \langle
+y|-y\rangle =0,
\end{equation}%
and order the basis as $\{|\!+\!y\rangle ,|\!-\!y\rangle \}$ (descending $%
m_{y}$). Define the rotated lowering operator 
\begin{equation}
K\equiv \eta ^{-}e^{i\frac{\pi }{2}\eta ^{x}}.  \label{eq:K_def_app}
\end{equation}%
We claim that $K$ has the strictly upper-triangular form 
\begin{equation}
K^{(y)}=%
\begin{pmatrix}
0 & -i\frac{\sqrt{2}}{2} \\ 
0 & i\frac{\sqrt{2}}{2}%
\end{pmatrix}%
\quad \text{in the ordered basis }\{|\!+\!y\rangle ,|\!-\!y\rangle \}.
\label{eq:K_upper_app}
\end{equation}

To prove this, it suffices to show that $K$ annihilates the $+y$ eigenstate.
Let $U\equiv e^{-i\frac{\pi}{2}\eta^x}$ so that $K=U^\dagger \eta^- U$. Then 
$K|\!+\!y\rangle=0$ is equivalent to $\eta^- U|\!+\!y\rangle=0$. Since $%
\eta^-$ annihilates the lowest-weight state in the $z$-quantization, the
condition $\eta^-|\phi\rangle=0$ implies $|\phi\rangle\propto|\downarrow_z%
\rangle$. But $U$ is precisely the rotation mapping $|\!+\!y\rangle$ to $%
|\downarrow_z\rangle$, hence 
\begin{equation}
K|\!+\!y\rangle = \eta^- e^{i\frac{\pi}{2}\eta^x}|\!+\!y\rangle = 0.
\label{eq:K_annihilates_plusy_app}
\end{equation}
Equation~\eqref{eq:K_annihilates_plusy_app} implies that the first column of 
$K^{(y)}$ vanishes: 
\begin{equation}
\langle +y|K|+y\rangle=0, \qquad \langle -y|K|+y\rangle=0,
\end{equation}
which proves Eq.~\eqref{eq:K_upper_app}. Therefore $K$ is strictly upper
triangular in the $\eta^y$ basis.

\subsection{Doubled-space ordering and triangularity of the jump term}

We now consider the doubled-space basis used in the main text, where each $%
\eta_j^y$ and $\tilde{\eta}_j^y$ is diagonal. A convenient product basis is
labeled by the local eigenvalues 
\begin{equation}
\eta_j^y|m_j\rangle = m_j|m_j\rangle,\qquad \tilde{\eta}_j^y|\tilde
m_j\rangle = \tilde m_j|\tilde m_j\rangle, \qquad m_j,\tilde m_j=\pm\frac12.
\end{equation}
We define the doubled basis vectors as 
\begin{equation}
|\{m_j,\tilde m_j\}\rangle\rangle \equiv
\bigotimes_{j=1}^{N}|m_j\rangle\otimes|\tilde m_j\rangle,
\end{equation}
and order them lexicographically with $m_j$ and $\tilde m_j$ both ordered as 
$(+\frac12)\succ(-\frac12)$ at each site.

Within this ordering, the diagonal operator 
\begin{equation}
D_j\equiv \frac12\left(\eta_j^{y}-\tilde{\eta}_j^{y}\right)
\end{equation}
is manifestly diagonal, while the ket--bra coupling term in Eq.~%
\eqref{eq:exact_liouvillian_form} is 
\begin{equation}
J_j\equiv \eta_j^{-}e^{i\frac{\pi}{2}\eta_j^{x}}\, \tilde{\eta}_j^{-}e^{-i%
\frac{\pi}{2}\tilde{\eta}_j^{x}} = K_j\,\tilde K_j,  \label{eq:Jj_def_app}
\end{equation}
where $K_j$ and $\tilde K_j$ act on the ket and bra degrees of freedom at
site $j$, respectively. By the single-site lemma above, both $K_j$ and $%
\tilde K_j$ are strictly upper triangular in their local $\eta^y$
eigenbases. Therefore the product $J_j=K_j\tilde K_j$ is also strictly upper
triangular on the local doubled space at site $j$, and hence cannot
contribute to any diagonal matrix element in the global doubled basis.

It follows that the full operator $\hat{\mathcal{L}}_{\mathrm{eff}}^{(0)}$
in Eq.~\eqref{eq:exact_liouvillian_form}, being a sum of diagonal terms and
strictly upper-triangular terms (in the above ordering), is itself strictly
upper triangular up to the diagonal part. In particular, the eigenvalues of $%
\hat{\mathcal{L}}_{\mathrm{eff}}^{(0)}$ are exactly given by its diagonal
entries, i.e., 
\begin{eqnarray}
\lambda _{\{m_{j},\tilde{m}_{j}\}} &=&\sum_{j\in \mathcal{J}}\left[ -\frac{%
\gamma }{2}+\gamma \,\langle \{m,\tilde{m}\}|D_{j}|\{m,\tilde{m}\}\rangle %
\right], \\
&=&\sum_{j\in \mathcal{J}}\left[ -\frac{\gamma }{2}+\frac{\gamma }{2}(m_{j}-%
\tilde{m}_{j})\right] ,
\end{eqnarray}%
which is Eq.~\eqref{eq:gap_main} of the main text.

\subsection{Minimal explicit example: one driven site}

For completeness, consider one driven site (drop the site index) and the
local doubled basis $\{|+y\rangle \otimes |+y\rangle ,\ |+y\rangle \otimes
|-y\rangle ,\ |-y\rangle \otimes |+y\rangle ,\ |-y\rangle \otimes |-y\rangle
\}$. In this basis, the diagonal contribution is 
\begin{equation}
-\frac{\gamma }{2}+\frac{\gamma }{2}\left( \eta ^{y}-\tilde{\eta}^{y}\right)
=\mathrm{diag}\left( -\frac{\gamma }{2},\ 0,\ -\gamma ,\ -\frac{\gamma }{2}%
\right) ,
\end{equation}%
while the jump term $J=K\tilde{K}$ is strictly upper triangular and thus
leaves the diagonal entries unchanged. Hence the diagonal part fully
determines the eigenvalues, and the smallest nonzero decay rate is $\gamma
/2 $, establishing the Liouvillian gap $\Delta =\gamma /2$.

\bibliographystyle{apsrev4-2}
\bibliography{SPOS_refs}

\end{document}